\documentclass[conference]{IEEEtran}
\IEEEoverridecommandlockouts
\usepackage{cite}
\usepackage{amsmath,amssymb,amsfonts}
\usepackage{algorithmic}
\usepackage{graphicx}
\usepackage{textcomp}
\usepackage{xcolor}
\usepackage{algorithm}
\usepackage{stfloats}
\usepackage{makecell}
\usepackage{enumitem}
\usepackage{xspace}
\usepackage{boldline}
\usepackage{arydshln}
\usepackage{hyperref}
\usepackage{multirow}
\usepackage{amsmath}
\usepackage{balance}
\usepackage[misc,geometry]{ifsym}
\def\BibTeX{{\rm B\kern-.05em{\sc i\kern-.025em b}\kern-.08em
    T\kern-.1667em\lower.7ex\hbox{E}\kern-.125emX}}

\newcommand\fullname{\emph{\underline{T}wo-\underline{S}tep \underline{P}rediction \underline{N}etwork with \underline{R}emote Sensing \underline{A}ugmentation}}
\newcommand\mymodel{TSPN-RA}
\newcommand\qtree{{\it quad-tree}\xspace}
\newcommand\qrpgraph{{\it QR-P graph}\xspace}

\begin{document}

\title{Towards Effective Next POI Prediction: Spatial and Semantic Augmentation with Remote Sensing Data\\
}

\author{\thanks{\Letter\ Corresponding author.}
\IEEEauthorblockN{Nan Jiang\IEEEauthorrefmark{2}, Haitao Yuan\IEEEauthorrefmark{3}\href{haitao.yuan@ntu.edu.sg}{\Letter}, Jianing Si\IEEEauthorrefmark{2}, Minxiao Chen\IEEEauthorrefmark{2}, Shangguang Wang\IEEEauthorrefmark{2}}
\IEEEauthorblockA{\IEEEauthorrefmark{2}Beijing University of Posts and Telecommunications, China  
\IEEEauthorrefmark{3}Nanyang Technological University, Singapore \\
\{jn\_bupt, sijianing, chenminxiao, sgwang\}@bupt.edu.cn, haitao.yuan@ntu.edu.sg}
}

\vspace{-0.5in}
\maketitle

\begin{abstract}
The next point-of-interest (POI) prediction is a significant task in location-based services, yet its complexity arises from the consolidation of spatial and semantic intent. This fusion is subject to the influences of historical preferences, prevailing location, and environmental factors, thereby posing significant challenges. In addition, the uneven POI distribution further complicates the next POI prediction procedure. To address these challenges, we enrich input features and propose an effective deep-learning method within a two-step prediction framework. Our method first incorporates remote sensing data, capturing pivotal environmental context to enhance input features regarding both location and semantics. 
Subsequently, we employ a region quad-tree structure to integrate urban remote sensing, road network, and POI distribution spaces, aiming to devise a more coherent graph representation method for urban spatial. Leveraging this method, we construct the QR-P graph for the user's historical trajectories to encapsulate historical travel knowledge, thereby augmenting input features with comprehensive spatial and semantic insights. 
We devise distinct embedding modules to encode these features and employ an attention mechanism to fuse diverse encodings. In the two-step prediction procedure, we initially identify potential spatial zones by predicting user-preferred tiles, followed by pinpointing specific POIs of a designated type within the projected tiles. Empirical findings from four real-world location-based social network datasets underscore the remarkable superiority of our proposed approach over competitive baseline methods.
\end{abstract}


\section{Introduction}
\label{sec:1}

In recent years, location-based services have experienced significant development, enabling users to conveniently share their checked-in locations or point-of-interest (POIs) in order to obtain better services. One of the most significant tasks is to utilize users' check-in history to predict their subsequent visits to POIs, a practice known as the next POI prediction. 
In practical applications, POI prediction models can offer user with real-time next destination recommendations in navigation software, based on their check-in records. They can also be utilized in travel apps to automatically plan suitable travel routes for users... Such models have the potential to enhance user experience across various scenarios.

Existing works on this task have explored various solutions. Early works mainly focus on seeking solutions with traditional methods, such as Markov Chains~\cite{gambs2012next,chen2014nlpmm} and Matrix Factorization~\cite{koren2009matrix}. Later, deep learning methods~\cite{elman1990finding}(i.e., RNN variants such as LSTM and GRU) have been utilised by several works~\cite{liu2016predicting, feng2018deepmove, sun2020go, lim2021origin}. More recently, the research focus has shifted towards attention-based and graph-based solutions. The former use the attention layer as the primary way to aggregate historical information~\cite{SAE-NAD, feng2018deepmove, luo2021stan, li2021mgsan, lim2021origin, STiSAN}, while the latter pay more attention to deal with non-successive information from POI transitions, constructing graphs based on high dimensional relationships~\cite{rao2022graph, lim2022hierarchical, lim2020stp}. To boost expressive learning ability, some methods introduce new data formats, such as text~\cite{yao2017serm, chang2018content,li2021tell} or POI categories~\cite{he2016inferring, lai2021spent+, li2021exploring, liu2022real, zhao2021hierarchical}. Aiming to cope with the sparsity problem, some studies~\cite{feng2018deepmove,lim2022hierarchical} partition spatial areas with hierarchical grids. 

\label{section:intro}
\begin{figure}
  \centering
  \includegraphics[width=\linewidth]{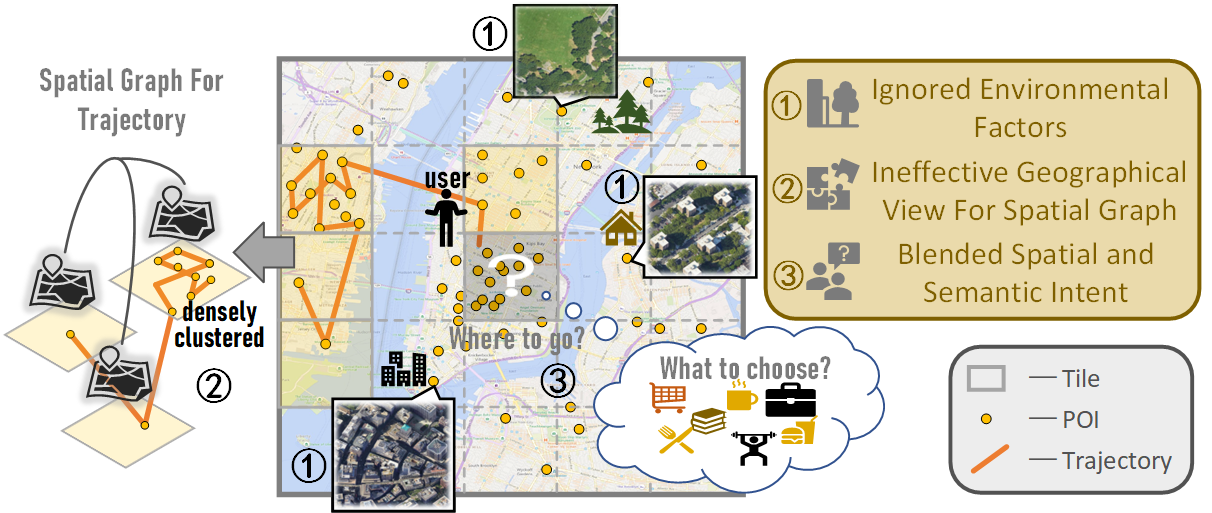}
  \vspace{-0.25in}
  \caption{Three key challenges for predicting the next POI.}
  \vspace{-0.2in}
  \label{fig:challenges}
\end{figure}

However, as shown in~\autoref{fig:challenges}, it remains an open problem to design an effective method to solve the problem for three reasons:

\noindent (1) \textbf{Neglecting Environmental Influences.} Traditional approaches\cite{feng2018deepmove,sun2020go,luo2021stan,lim2022hierarchical,rao2022graph} often fail to consider crucial environmental factors like road density and land usage patterns, which significantly impact the visitation patterns of POIs. However, harnessing these factors can empower systems to gain a deeper comprehension of urban dynamics and consequently improve the precision of predictions. For instance, when a specific POI is surrounded by high road density, it logically implies that it is more likely to attract visitors during peak commuting hours due to its accessible location.

\noindent (2) \textbf{Ineffective Geographical View for Spatial Graph.} There often exist non-uniform distributions of POIs, resulting in pronounced imbalances across different geographical areas. Conventional methods, like hierarchical grids~\cite{feng2018deepmove,lim2022hierarchical}, lead to partitioned areas with vastly differing POI densities. Therefore, when constructing a spatial graph for trajectories, this uneven distribution can cause POIs to densely cluster within the same tile nodes, thereby reducing the spatial representation capability of the graph. Moreover, existing methods ignore the connectivity of road networks between different regional partitions, resulting in a lack of modeling spatial constraints.

\noindent (3) \textbf{Blended Spatial and Semantic Intents.} The next POI prediction encompasses two distinct yet interconnected dimensions: spatial prediction and trajectory semantic prediction. Spatial prediction entails identifying the general area where the user might visit next, while trajectory semantic prediction involves predicting the specific type of place they are likely to choose within that area. However, current methods\cite{luo2021stan,lim2022hierarchical,rao2022graph} in a unified prediction framework often fail to explicitly capture the correlation between these two intentions, leading to ambiguity and imprecision.

To this end, we present our innovative solution, named \fullname\ (\mymodel), for the next POI prediction task. At first, we \textbf{take a significant step by incorporating remote sensing imagery to address the first challenge}. This data augmentation infuses crucial environmental context into the prediction process, enriching the understanding of the areas surrounding POIs.
Next, we \textbf{involved the quad-tree structure when constructing a spatial knowledge graph in response to the second challenge}. The partitioning method allows us to adjust the observation granularity according to the density of POIs, In particular, we further enhance the tree-like structure with the road network and construct a graph, denoted as \qrpgraph, to capture the spatial constraint between different remote sensing tiles and their contained POI nodes. 
Finally, \textbf{in response to the third challenge, we design \mymodel\ with a two-step prediction framework}. This framework stands as a pivotal solution to effectively address the intricate relationship between spatial and semantic intents. Firstly, the model identifies potential tiles for user visits, effectively addressing spatial intent. Secondly, it leverages this spatial understanding to recommend the most relevant POIs within these tiles, thus fulfilling semantic intent. To elaborate, we start by extracting tile and POI features from historical and current prefix trajectories, which are then encoded using respective embedding modules. To capture spatio-temporal correlations, we enhance these embeddings with spatial and temporal encoders. Furthermore, we integrate user's historical trajectories with the quad-tree and road network to form a historical preference knowledge graph, referred to as \qrpgraph. This graph is used with an attention mechanism to inject historical preference into tile and POI feature encodings. Finally, we select the top-$K$ candidate tiles based on tile feature encodings and predict the next POI among POI candidates in these tiles using POI feature encodings. Leveraging the two-step prediction framework, we can offer personalized POI recommendations that align closely with users' spatial and semantic preferences, significantly enhancing the user experience across various practical applications.

In summary, the contributions of this paper are as follows: 
\begin{itemize}[leftmargin=10.2pt]
\setlength{\itemsep}{0pt}
\setlength{\parsep}{0pt}
\setlength{\parskip}{0pt}
    \item We propose \mymodel, a two-step prediction network, predicting the next POI at both the spatial and semantic levels. Our model employs a more effective strategy for partitioning urban areas and leverages environmental factors to enhance spatial and semantic representations.(Sec.~\ref{sec:2}\&~\ref{sec:3})
    \item We incorporate remote sensing images to embed tile areas, enriching them with environmental factors. We enhance embeddings by incorporating spatial and temporal information. Furthermore, we introduce the heterogeneous graph \qrpgraph to capture historical preference knowledge, and apply the graph neural network to learn its embedding.(Sec.~\ref{sec:4})
    \item We utilize the attention mechanism to fuse diverse embeddings, followed by a spatial and semantic two-step prediction process, where the first step is to select candidate tiles, and the second step is to rank POIs in all candidate tiles.(Sec.~\ref{sec:5})
    \item We perform a thorough evaluation on four real-world datasets. The results show that \mymodel\ outperforms all the competitive state-of-the-art models in terms of both effectiveness and efficiency(Sec.~\ref{sec:6})

\end{itemize}

\section{preliminaries}
\label{sec:2}

\begin{figure}
  \centering
  \includegraphics[width=0.75\linewidth]{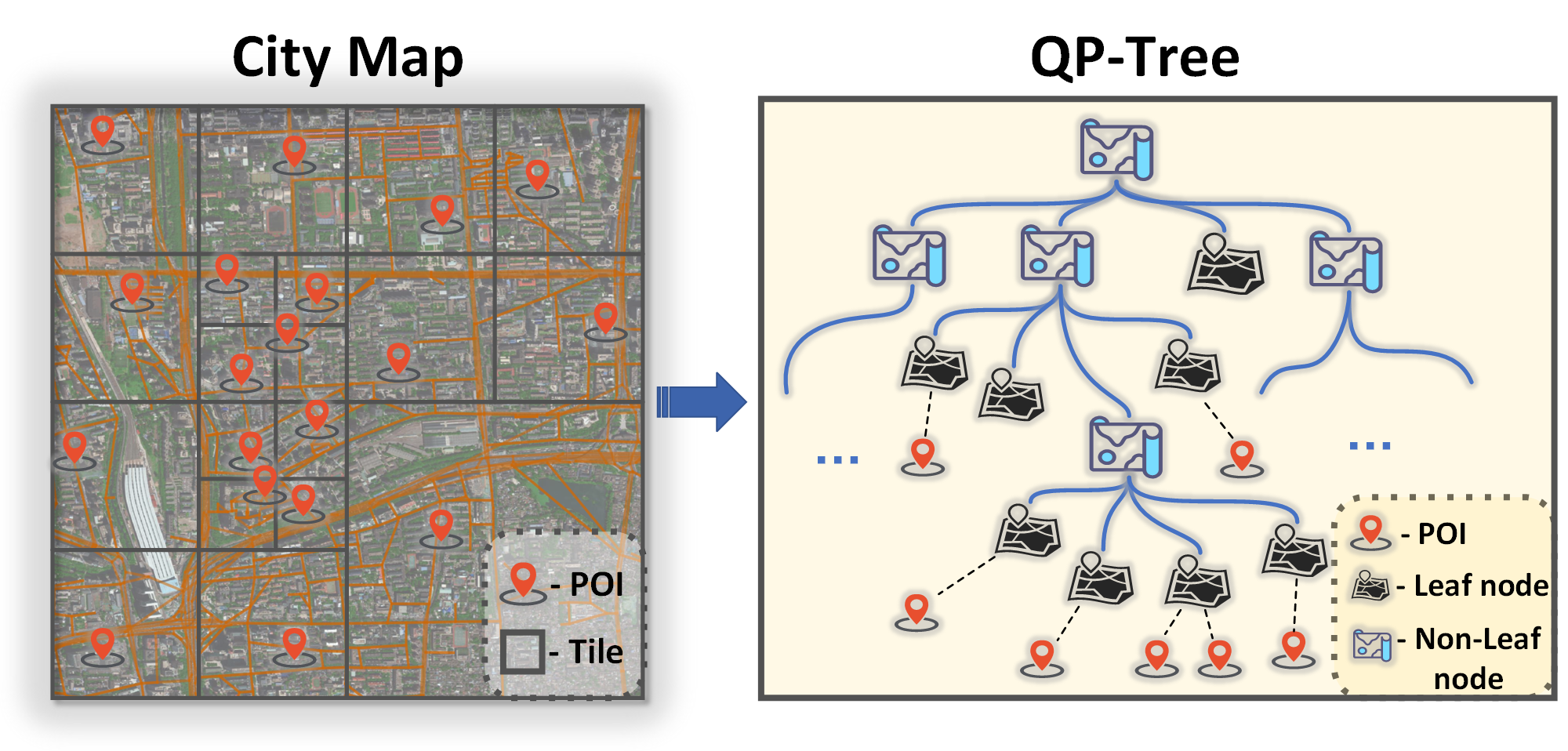}
  \vspace{-0.2in}
  \caption{An illustration of how to form a \qtree. A tile is divided into sub-tiles if there are too many POIs located in this region.}
  \label{fig:quad-tree}
  \vspace{-0.2in}
\end{figure}

In this section, we first introduce some key concepts and then formalize the next POI prediction problem. 

\subsection{Basic Concepts}
\label{sec:2.1}
\noindent \textbf{POI.} Each POI is defined as a tuple $p=(id, loc, cate)$, where $id$ means the identification number, $loc$ denotes the corresponding geographical coordinates (i.e., latitude and longitude), and $cate$ means the POI's category. In addition, all POIs are denoted as the set $P=\{p_1,\cdots,p_{|P|}\}$. 
\vspace{0.2em}

\noindent \textbf{Trajectory.} Given a time window $T=[t_{\alpha}, t_{\beta})$, we define the trajectory of a user $u$ as a sequence of the user's POI visiting records $S_{T}=[(p_1,t_1), (p_2, t_2), \cdots,  (p_{|S|},t_{|S|})]$ during the time window, where each record is composed of a POI $p_i$ and its corresponding visited timestamp $t_i$ (i.e., $t_{\alpha}\leq t_i< t_{\beta}$). Similar to~\cite{feng2018deepmove}, we generate disjoint time windows (i.e., $T_1=[t_{\alpha_1},t_{\beta_1}],T_2=[t_{\alpha_2},t_{\beta_2}],\cdots$) by incorporating an acceptable time gap $\Delta t$ (72 hours in this paper), where $t_{\alpha_{i+1}}-t_{\beta_i}\geq \Delta t$ is valid for any two adjacent windows. Therefore, we can generate a trajectory sequence $\mathcal{S}^u=[S^u_{T_1},S^u_{T_2},\cdots]$ for the user $u$. In particular, given the current trajectory $S^u_{T_i}$, we denote its prefix trajectories in the sequence $\mathcal{S}^u$ as historical trajectories $\mathcal{S}^u_{\vartriangleleft i}=[S^u_{T_1},\cdots, S^u_{T_{i-1}}]$.

\noindent \textbf{Region Quad-tree.} We incorporate the basic region \qtree ~\cite{finkel1974quad, wang2021adq} structure, to manage all POIs. As shown in Figure~\ref{fig:quad-tree}, the \qtree begins by initializing the root node, which corresponds to the entire considered region. When the POIs within a node's region surpass a predefined threshold, the structure recursively subdivides the space into smaller rectangular cells, creating a hierarchical tree of nodes. These cells, referred to as "tiles" in subsequent discussions, are integral components of the \qtree. Particularly, we denote the \qtree as $\mathbb{Q}$, and each node $v \in \mathbb{Q}$ pertains exclusively to either a leaf node or a non-leaf node. Additionally, all the leaf tiles, when stitched together, can completely cover the entire considered region, where any POI can be uniquely positioned within a specific leaf tile. 

\noindent \textit{Discussion.} The rationale behind the adoption of the quad-tree structure for spatial partitioning is twofold. Firstly, in contrast to alternative architectural frameworks (such as the grid index~\cite{lim2022hierarchical}), the quad-tree incorporates tiles with varying degrees of granularity, thereby facilitating the identification and analysis of multi-scale spatial correlations among disparate POIs. Secondly, the distribution of POIs in the quad-tree structure ensures a uniform dispersion across distinct leaf tiles, thereby mitigating the propensity to disproportionately accentuate regions devoid of POIs.

\subsection{QR-P graph} 
\label{section: QR-P graph}

\begin{figure}
  \centering
  \includegraphics[width=\linewidth]{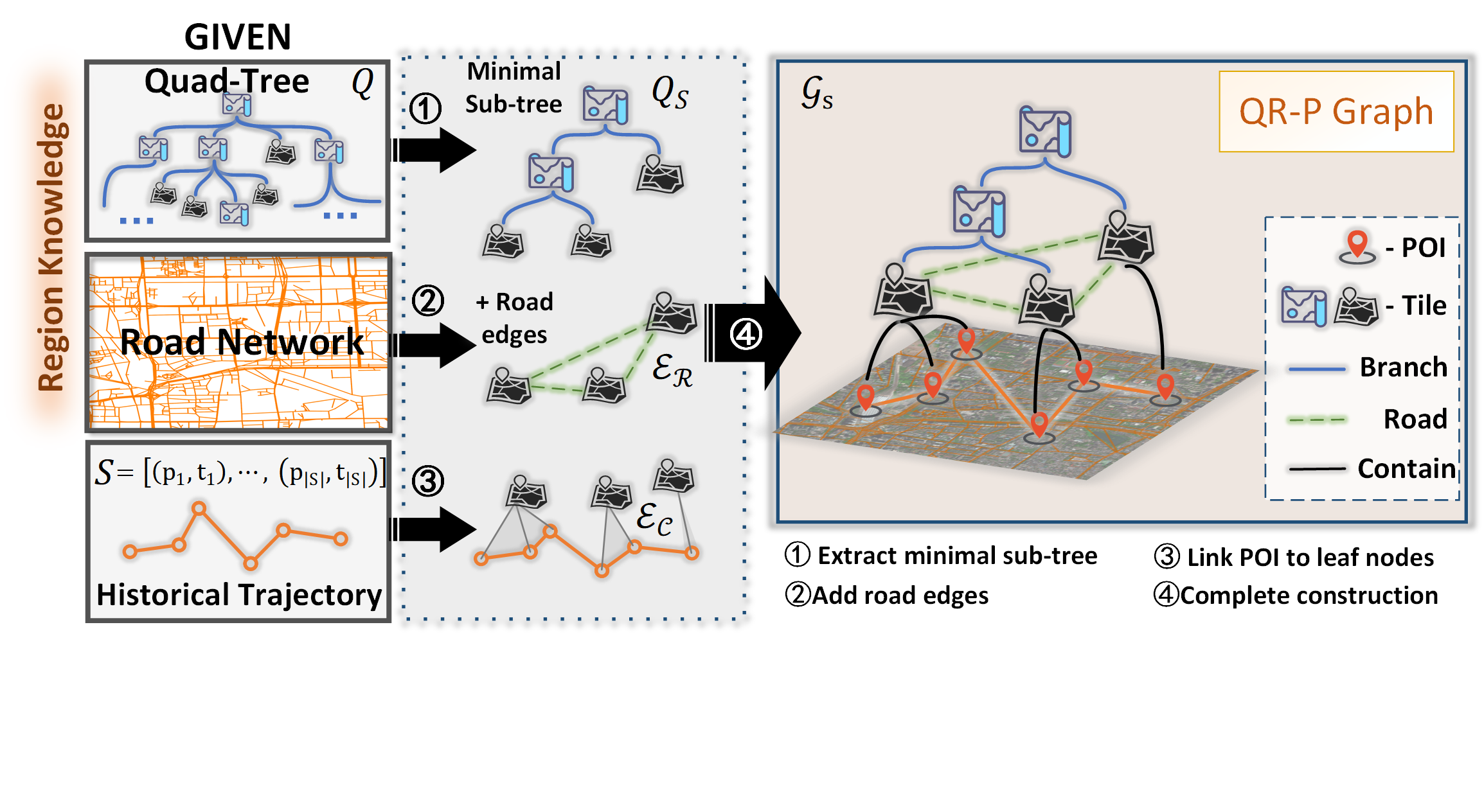}
  \vspace{-0.7in}
  \caption{The construction of \qrpgraph using quad-tree,  road network and historical trajectory. Detail definition is illustrated in \autoref{section: QR-P graph}.}
  \label{fig:QR-P graph}
  \vspace{-0.2in}
\end{figure}

Quad-tree can serve as a means of region representation; however, for certain adjacent small tiles near large tile boundaries, situations may arise where they are adjacent but face difficulty in exchanging information with each other. To address this limitation, we propose a strategic intervention by leveraging the road network to convert the tree structure into a more interconnected graph. Consequently, in order to capture localized correlations among distinct POIs within a trajectory, we integrate all these aspects and advance the concept of the QR-P(\underline{Quad-tree, Road network with Poi}) graph. The following introduces its detailed construction process.

\noindent \underline{\textit{Construction:}} To construct \qrpgraph for a given user trajectory $\text{\small$S=[(p_1,t_1),\cdots, (p_{|S|},t_{|S|})]$}$, we need the region \qtree $\mathbb{Q}$ and the region's road network. As illustrated in \autoref{fig:QR-P graph}, the construction procedure includes the following four steps:

{\fontsize{10}{10.5}\selectfont
\noindent (1) At first, we extract the minimal sub-tree from $\mathbb{Q}$ for the given trajectory $S$, denoted as $\mathbb{Q}_S$. This sub-tree's leaf nodes encompass all POIs within $S$, ensuring no smaller sub-tree could achieve the same coverage.

\noindent (2) Next, we derive the adjacency correlations among all leaf nodes within $\mathbb{Q}_S$ based on the road network. Specifically, for any pair of leaf nodes $\langle v_1, v_2 \rangle$ in $\mathbb{Q}_S$, we establish an edge from $v_1$ to $v_2$ if there's a direct road network link between the tiles respectively corresponding to $v_1$ and $v_2$. For simplicity, we denote these generated edges as road edges $\mathcal{E}_R$.

\noindent (3) In addition, we consider each POI in $S$ as a POI node, which is linked to its respective tile node via a containment-type edge, establishing their spatial relationship within the graph. For simplicity, we denote these containment-type edges as contain edges $\mathcal{E}_C$.

\noindent (4) Finally, we integrate $S$, $\mathbb{Q}_S$, $\mathcal{E}_R$, and $\mathcal{E}_C$ to form the \qrpgraph.
}

In summary, the \qrpgraph comprises two types of nodes (i.e., POI and \textit{tile}) and three categories of edges (i.e., \textit{branch}, \textit{road}, and \textit{contain}). Therefore, we can define this graph structure formally as follows:

\noindent \underline{\textit{Definition~[\qrpgraph]:}} Given a \qtree $\mathbb{Q}$, road network, and user trajectory $\text{\small$S=[(p_1,t_1),\cdots, (p_{|S|},t_{|S|})]$}$, we extract the minimal sub-tree $\mathbb{Q}_S$, the road edges $\mathcal{E}_R$ and the contain edges  $\mathcal{E}_C$. Next, we integrate $S, \mathbb{Q}_S, \mathcal{E}_R, \mathcal{E}_C$ as the \qrpgraph 
$\mathcal{G}_S=\left \langle \mathcal{V}_S,\mathcal{E}_S,\Phi_S,\Psi_S\right \rangle$, where:
{\fontsize{10}{10.5}\selectfont
\begin{itemize}[leftmargin=10.2pt]
\setlength{\itemsep}{0pt}
\setlength{\parsep}{0pt}
\setlength{\parskip}{0pt}
    \item $\mathcal{V}_S$ is the vertice set, associated with a type mapping function \( \Phi_S: \mathcal{V}_S \rightarrow T_V \), which assigns each node to a type \( t \in T_V \), where \( T_V = \{ POI, tile \} \). For vertice $v \in \mathcal{V}_S$:
    \begin{itemize}
        \item \( \Phi_S(v) = POI\): if $v\in $ $\{p_1,\cdots,p_{|S|}\}$.
        \item \( \Phi_S(v) = tile\): if $v$ represents a tree node in $\mathbb{Q}_S$.
    \end{itemize}
    \item $\mathcal{E}_S$ is the edge set, associated with a type mapping function \( \Psi_S: \mathcal{E}_S \rightarrow T_E \), which assigns each edge to a type \( t \in T_E \), where \( T_E = \{ branch, road, contain \} \). For edge $e \in \mathcal{E}_S$:
    \begin{itemize}
        \item \( \Psi_S(e) = branch\): if $e$ represents a tree edge in $\mathbb{Q}_S$, which connects two $tile$ type vertices that have the parent-children relationship.
        \item \( \Psi_S(e) = road\): if $e \in \mathcal{E}_R$, which connects two $tile$ type vertices that are both leaf nodes in $\mathbb{Q}_S$.
        \item \( \Psi_S(e) = contain\): if $e\in \mathcal{E}_C$, which connects a $tile$ type vertice with a $POI$ type vertice.
    \end{itemize}
\end{itemize}
}

\subsection{Remote Sensing Augmentation\label{section:rs}} 
\begin{figure}
    \centering
    \begin{minipage}{0.4\linewidth}
    \centering
    \includegraphics[width=\linewidth]{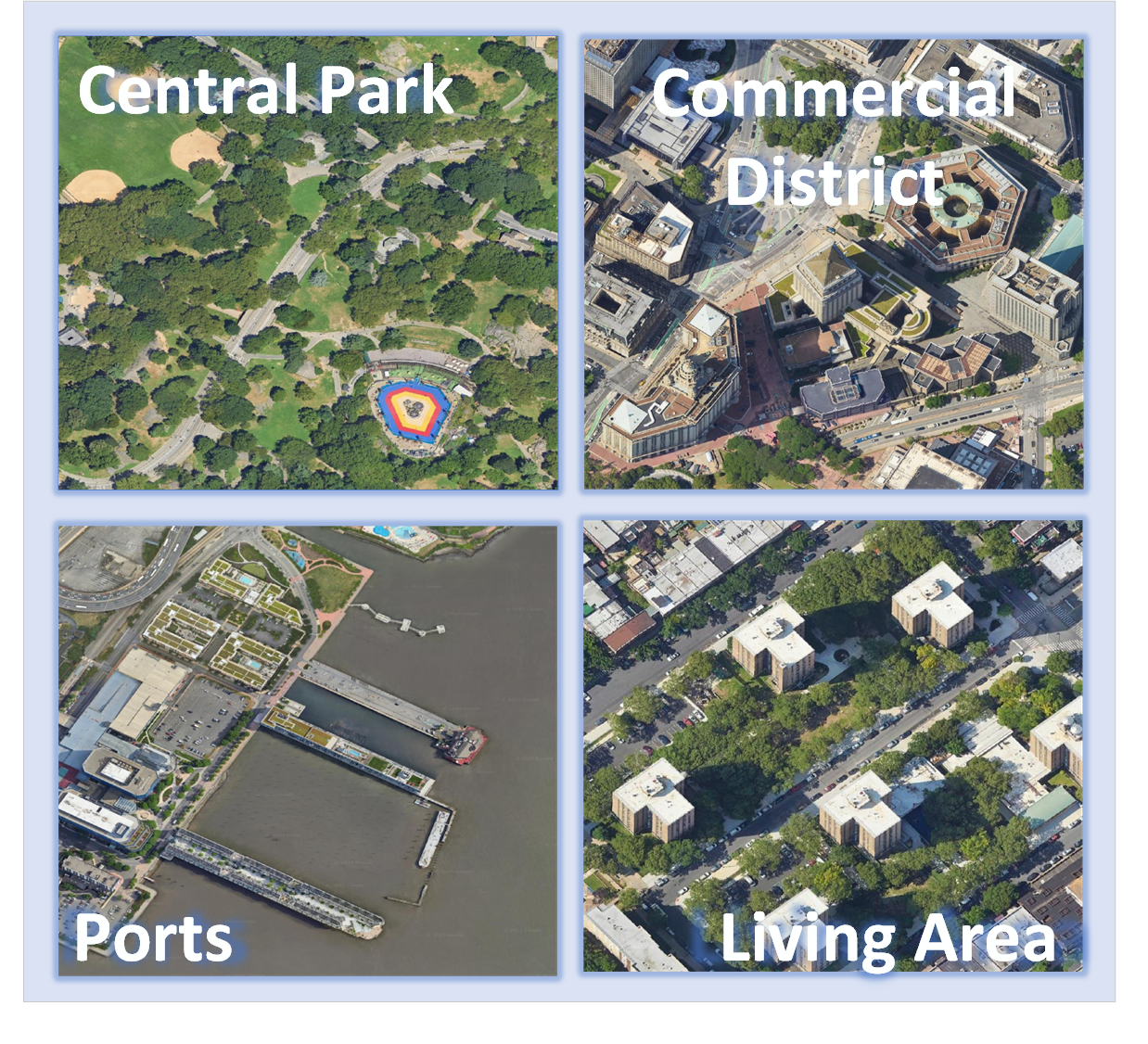}
    \end{minipage}%
    \begin{minipage}{0.4\linewidth}
    \centering
    \includegraphics[width=\linewidth]{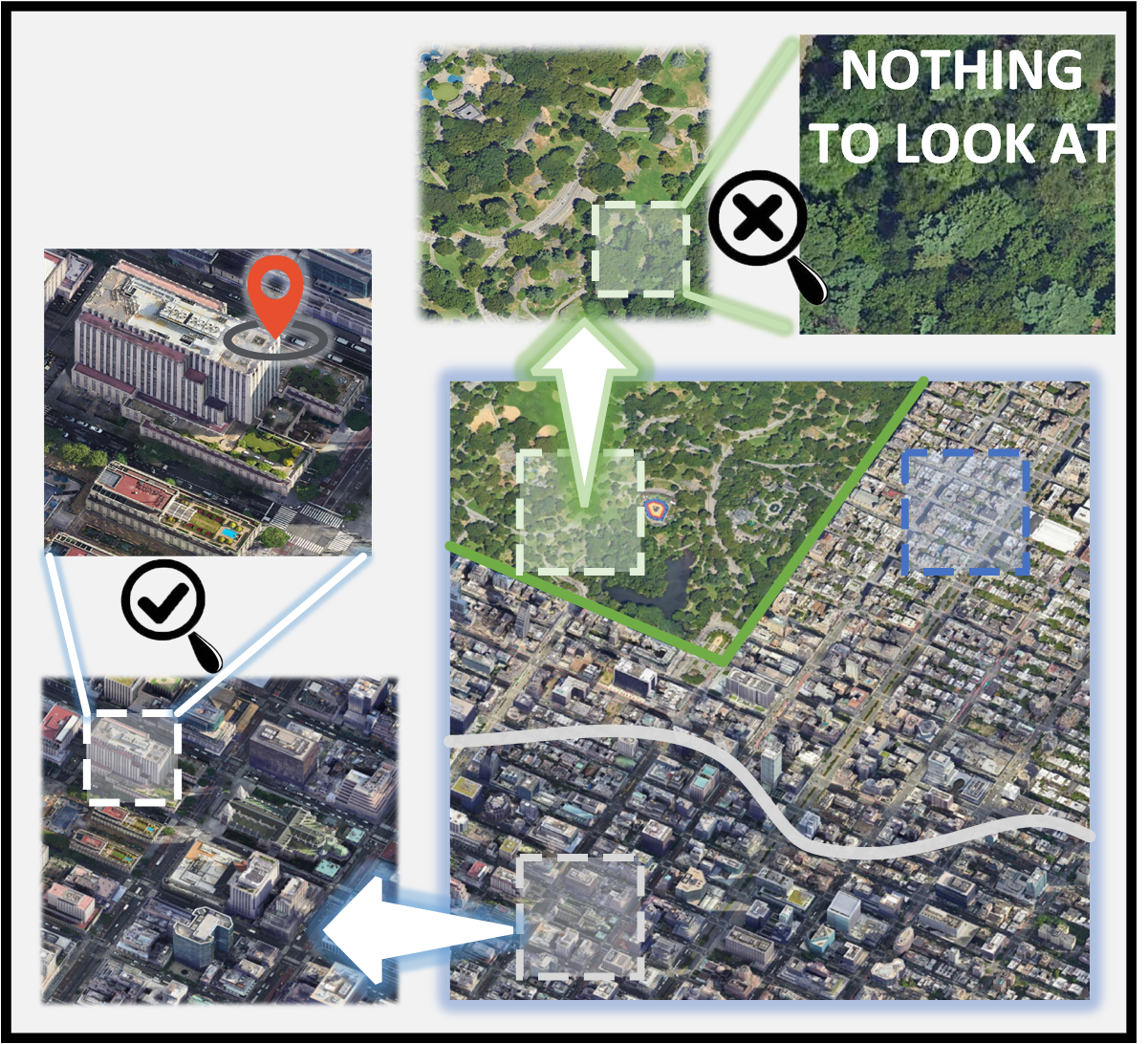}
    \end{minipage}
    \caption{The left part shows the aerial view of different kinds of districts. The right part illustrates that large-scale low-resolution imagery shows a clear view of area distribution, while small-scale high-resolution imagery contains detailed environmental information for POI neighborhood. However, it is unnecessary to get high-resolution vision of some areas, such as parks, that have repetitive views without much human mobility.}
    \label{figure: remote sensing}
    \vspace{-0.2in}
\end{figure}

Remote sensing includes various types of data\footnote{LiDAR, SAR, Thermal imagery, RGB/Hyperspectral imagery}, among which satellite imagery is the aerial view of the ground photoed by low-Earth orbit remote sensing satellite. Nowadays, the resources of well-processed (clear and cloud-free) RGB satellite imagery are abundant, we can easily obtain satellite imagery of specific area from Google Map\footnote{https://www.google.com/maps}. As shown in the left part of \autoref{figure: remote sensing}, we can observe that it is not hard to distinguish the difference between different kind of districts by the shape and colors. 
Moreover, the right part of \autoref{figure: remote sensing} reveals satellite imagery in different scales. In large-scale imagery, we can view the city districts and the distribution of different urban environments. While in small-scale imagery, we can get a clear view of the details of the neighborhood. This gives us the intuition that satellite imagery can serve as a warehouse for spatial representation learning in urban regions.

\subsection{Problem Formulation}
\label{sec:problem}

Given a specific time window $T_i$, a user's historical trajectories $\mathcal{S}_{\vartriangleleft i}=\{S_{T_1}, \cdots, S_{T_{i-1}}\}$, and the prefix sequence $\{(p_1,t_1), (p_2,t_2), \cdots, $$(p_{j-1},t_{j-1})\}$ of the current trajectory $S_{T_i}$, the next POI recommendation task is to generated an ordered set of POIs, denoted as $R_P$, such that the actual visited POI $p_{j}$ at the subsequent time stamp $t_j$ holds a prominent rank within the set $R_P$. In particular, the objective function can be denoted as follows:
\begin{equation} 
\small
\begin{aligned}
    &\text{argmin}_{\theta}\  index(p_j, R_P) \\
    &s.t., R_P =\mathcal{M}_{\theta}(\mathcal{S}_{\vartriangleleft i},S_{T_i}\text{[1:j-1]})
\end{aligned}
\end{equation} 
where $index(p_j, R_P)$ corresponds to locating the index of $p_j$ within the array $R_P$ (if $p_j$ is not present in $R_P$, we would have $index(p_j, R_P) = |R_P| + 1$), $\mathcal{M}_{\theta}(\cdot)$ is a learned model based on the given inputs.

\section{model overview}
\label{sec:3}
\begin{figure}
  \centering
  \includegraphics[width=0.9\linewidth]{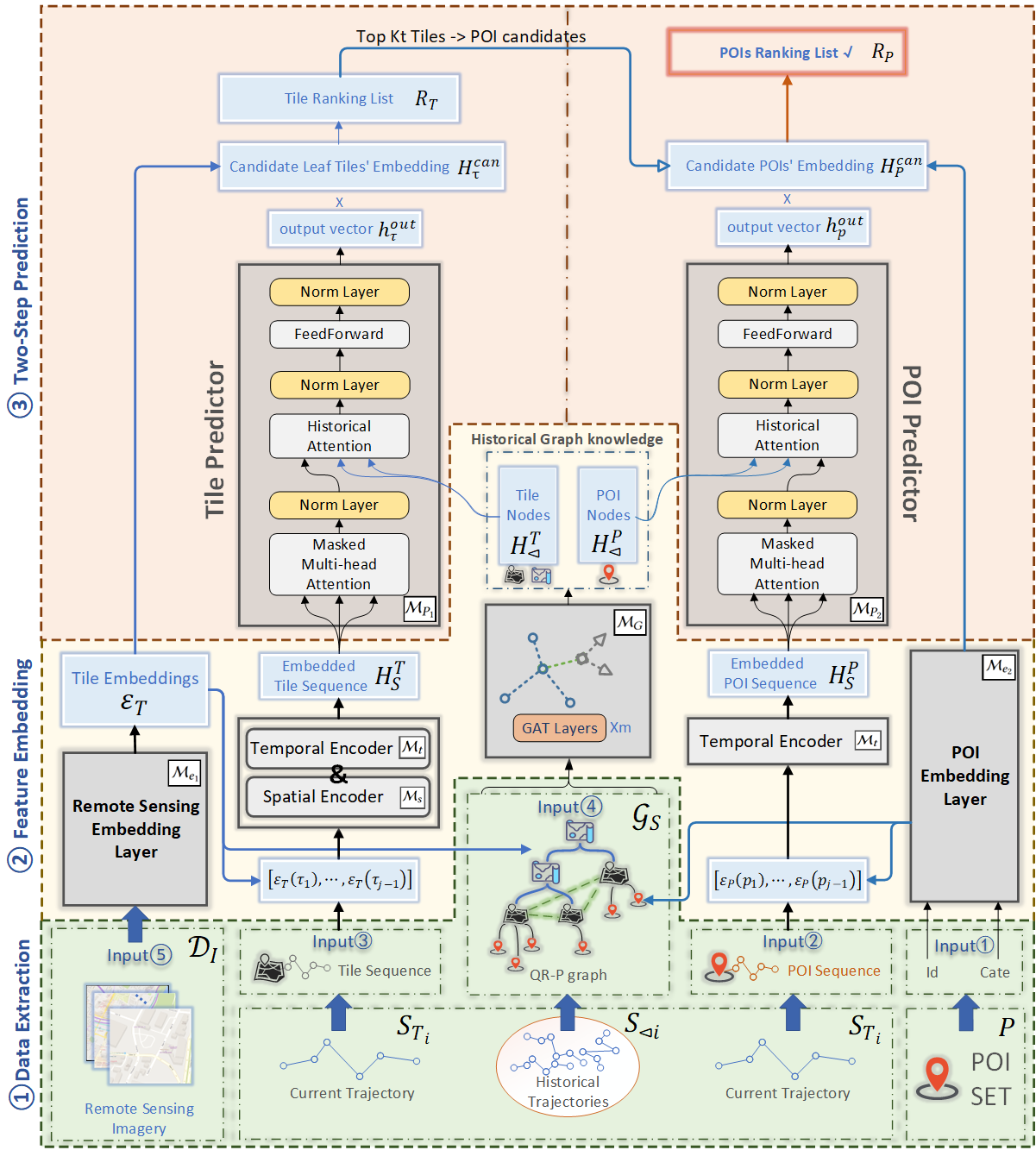}
  \caption{The main architecture of \mymodel. The whole model can be separated into three main sections: Data Extraction, Feature Embedding, and Two-step Prediction. These sections are represented by green, yellow, and red backgrounds(the same goes for subsequent figures). Additionally, the main modules are drawn as squares with dark edges. Blue squares with light edges represent intermediate data.}
  \label{fig:model architecture}
  \vspace{-0.2in}
\end{figure}

In this section, we outline the architecture of our proposed \mymodel\ model. As shown in~\autoref{fig:model architecture}, the comprehensive framework consists of three parts: \textit{Data Extraction}, \textit{Feature Embedding}, and \textit{Two-Step Prediction}. In particular, we first use \textit{Data Extraction} to transform raw input into spatio-temporal features, such as POI and tile sequences. Next, we fed these features into \textit{Feature Embedding} to get different embeddings. Finally, a systematic two-step approach is devised within the \textit{Two-Step Prediction} part, encompassing the initial selection of candidate tiles in the first step, followed by the subsequent identification of optimal POIs within these chosen candidate tiles. Next, we will provide detailed components of the three parts in the subsequent subsections.

\noindent \underline{\textit{phase~1~[\textbf{Data Extraction.}]}}
As illustrated in the first section of Figure \ref{fig:model architecture}, the model requires five primary data sources as input. The following provides an explanation of these inputs.
\vspace{-0.5\topsep}
\begin{itemize}[leftmargin=10.2pt]
\setlength{\itemsep}{0pt}
\setlength{\parsep}{0pt}
\setlength{\parskip}{0pt}
\item \textbf{POI set}: We consider both the id and the category data for each POI in the POI set $P=\{p_1,\cdots,p_{|P|}\}$. 

\item  \textbf{POI sequence}: As mentioned in~\autoref{sec:problem}, we take the prefix POI sequence $S_{T_i}$[1:j-1] of the current trajectory as input.    

\item \textbf{Tile sequence}: To capture the spatial correlation in a coarse-grained view, we construct a sequence of leaf tiles based on the above POI sequence. In particular, we project each POI $p$ into the leaf tile $\tau$, and hence get the sequence $S_{T_i}^{tile}$[1:j-1]\\=$\{(\tau_1,loc_1,t_1),\cdots, (\tau_{j-1},loc_{j-1},t_{j-1})\}$, where $loc$ and $t$ indicate the specific location and timestamp, respectively. 

\item \textbf{QR-P Graph}: To better capture the complex spatio-temporal correlation among historical trajectories, we build the \qrpgraph according to the introduced procedure in~\autoref{section: QR-P graph}. In particular, we first concat the historical trajectories $\mathcal{S}_{\vartriangleleft i}=\{S_{T_1}, \cdots, S_{T_{i-1}}\}$ to a whole trajectory sequence according to the order of time. Then, the whole trajectory is used as the input sequence to generate QR-P graph $\mathcal{G}_S$. To this end, $\mathcal{G}_S$ takes the place of $\mathcal{S}_{\vartriangleleft i}$ to represent and learn historical information.

\noindent \textit{Discussion.} The motivation behind adopting the \qrpgraph lies in its capability to effectively replace historical trajectories. On the one hand, \qrpgraph is generated based on historical trajectories, so it involves historical travel patterns/knowledge. On the other hand, there exist intricate correlations between regions and the distribution of historically visited POIs, and the unique structure and features of \qrpgraph make it the most suitable candidate to encapsulate the complexity of these relationships.

\item \textbf{Tile set with remote sensing images}: As mentioned in~\autoref{section:rs}, we enhance each tile with its corresponding remote sensing image. Specifically, we use each tile's boundary box to extract a 256*256 sized image from Google Map. To this end, we can get all remote sensing images, which are denoted as $\mathcal{D}_{I}=\{I_1,I_2,\dots,I_{|\mathcal{D}_{I}|}\}$.  

\vspace{-.2em}
\end{itemize} 

\noindent \underline{\textit{phase~2~[\textbf{Feature Embedding.}]}}
This stage aims to embed extracted features by incorporating spatio-temporal correlations. 

\noindent \textbf{Embedding POIs and tiles}. We introduce two distinct modules, $\mathcal{M}_{e_1}$ and $\mathcal{M}_{e_2}$, each catered to embedding tile nodes and POIs, correspondingly. In the context of $\mathcal{M}_{e_1}$, a Convolutional Neural Network undertakes the task of fuse environmental information sourced from remote-sensing tile imagery dataset $\mathcal{D}_{I}=\{I_i,I_2,\dots,I_{|\mathcal{D}_{I}|}\}$. The resulting hyper-image is flattened and normalized, engendering a cluster of adaptable tile embeddings $\mathcal{E}_{T}$, with each tile's embedding represented as: $\mathcal{E}_{T}(\tau)=\mathcal{M}_{e_1}(I_{\tau})$. Similarly, POI embeddings $\mathcal{E}_{P}$ are formulated using $id$ and $cate$ information derived from POI tuples. Here, the embedding of a POI $p=(id, loc, cate)$ is rendered as: $\mathcal{E}_{P}(p)=\mathcal{M}_{e_2}(id,cate)$. 

\noindent \textbf{Embedding POI sequence and tile sequence.} Given a user's POI sequence $S$ and its corresponding tile sequence $S^{tile}$, we first fetch each POI/tile's embedding from $\mathcal{E}_{P}$/$\mathcal{E}_{T}$, and then respectively introduce some modules to further encode them by fusing sample-wise spatio-temporal information, including accurate locations and timestamps. In particular, the \textit{Spatial Encoder} module $\mathcal{M}_s$ is used to incorporate the corresponding POI location $loc_k$ for each tile $\tau_k$, while the \textit{Temporal Encoder} module $\mathcal{M}_t$ is used to fuse the specific timestamp $t_k$ for each POI  and its corresponding tile $\tau_k$. Therefore, we can generate the enhanced embeddings as follows:
\begin{equation}
\begin{small}
\begin{aligned}
    h_{\tau_k} &= \mathcal{M}_{t}(\mathcal{M}_{s}(\mathcal{E}_{T}(\tau_k),loc_k),t_k)\\
    h_{p_k} &= \mathcal{M}_{t}(\mathcal{E}_{P}(p_k),t_k)
    \end{aligned}
\end{small}
\end{equation}
As for the two sequences, we respectively obtain two embedding sequences $H^{P}_{S} = [h_{p_1},\dots,h_{p_{j-1}}]$ and $H^{T}_{S} = [h_{\tau_1},\dots,h_{\tau_{j-1}}]$. 

\noindent \textbf{Encoding QR-P graph to generate historical graph knowledge.}  We devise the module $\mathcal{M}_{G}$ to extract historical insights from the QR-P graph $\mathcal{G}_S=\left \langle \mathcal{V}_S,\mathcal{E}_S,\Phi_S,\Psi_S \right \rangle$. Considering that $\mathcal{G}_S$ is a heterogeneous graph, we first fetch initial embeddings from $\mathcal{E}_T$ and $\mathcal{E}_P$ for tile nodes and POI nodes, respectively.  Next, we apply a graph neural network to generate tile-level and POI-level historical knowledge, that are embedded in tile and POI node embeddings, namely $
    H^T_{\vartriangleleft}, H^T_{\vartriangleleft} = \mathcal{M}_{G}(\mathcal{G}_S, \mathcal{E}_T, \mathcal{E}_P)$.

\underline{\textit{phase~3~[\textbf{Two-Step Prediction.}]}}
Given the embeddings for tiles ($\mathcal{E}_{T}$) and POIs ($\mathcal{E}_{P}$), as well as the embedded sequences of tiles and POIs ($H^{T}_S$ and $H^{P}_S$) and historical graph knowledge ($H^T_{\vartriangleleft}$ and $H^P_{\vartriangleleft}$), two distinct prediction modules, denoted as $\mathcal{M}_{P_1}$ and $\mathcal{M}_{P_2}$, are employed to predict the most probable tiles and POIs for the forthcoming visit. These modules are constructed with an attention mechanism, allowing them to grasp self-attention patterns from the ongoing embedded sequence. Additionally, we incorporate a cross-attention element to consolidate insights from historical graph knowledge. This amalgamation results in the production of output vectors (referred to as $h_{\tau}^{out}$ and $h_{p}^{out}$) employed for the prediction. The process is outlined as follows:
\begin{equation}
\small
\begin{aligned}
h_{\tau}^{out} &= \mathcal{M}_{P_1}(H^T_S, H^T_{\vartriangleleft}) \\
h_{p}^{out} &= \mathcal{M}_{P_2}(H^{P}_S, H^P_{\vartriangleleft})
\end{aligned}
\small
\end{equation}
At last, we conduct the two-prediction process to generate results based on $h_{\tau}^{out}$ and $h_{p}^{out}$. 

\noindent \textbf{Tile prediction.} In the first prediction step, some candidate tiles are selected and their corresponding embeddings are fetched to form the set $H_\tau^{can} = \{\mathcal{E}_{T}(\tau)\ |\ \forall \mathcal{V}_\omega(\tau) = \emph{Leaf}\}$. Next, we rank these candidate tiles with $R_T=Sort(Sim(h_{\tau}^{out},H_\tau^{can}))$,
where \emph{Sort} is the descending sort function and \emph{Sim} represents the similarity measurement.

\noindent \textbf{POI prediction.} In the second prediction step, the top $K$ ranked tiles are selected to screen POI candidates, which is denoted as {\small$H_p^{can} = \{\mathcal{E}_{P}(p)|p\rightarrow\tau, \forall \ \tau\in \{R_T[1], \cdots, R_T[K]\}\}$}, where the symbol "$\rightarrow$" represents that POI $p$ is located in the boundary box of tile $\tau$. At last, the predicted POI ranking result $R_P$ can be denoted as:
$R_P=Sort(Sim(h_p^{out},H_p^{can}))$, where $R_P$ is the final output of \mymodel, as well as the required POI ranking list.

\section{Spatio-temporal Feature Embedding}
\label{sec:4}
In this section, we elaborate on the part of feature embedding, involving tile embedding, POI embedding, and the generation of historical graph knowledge based on encoding \qrpgraph.  

\subsection{Tile Embedding}
\label{section:tile}

\begin{figure}[!t]
  \centering
  \includegraphics[width=0.8\linewidth]{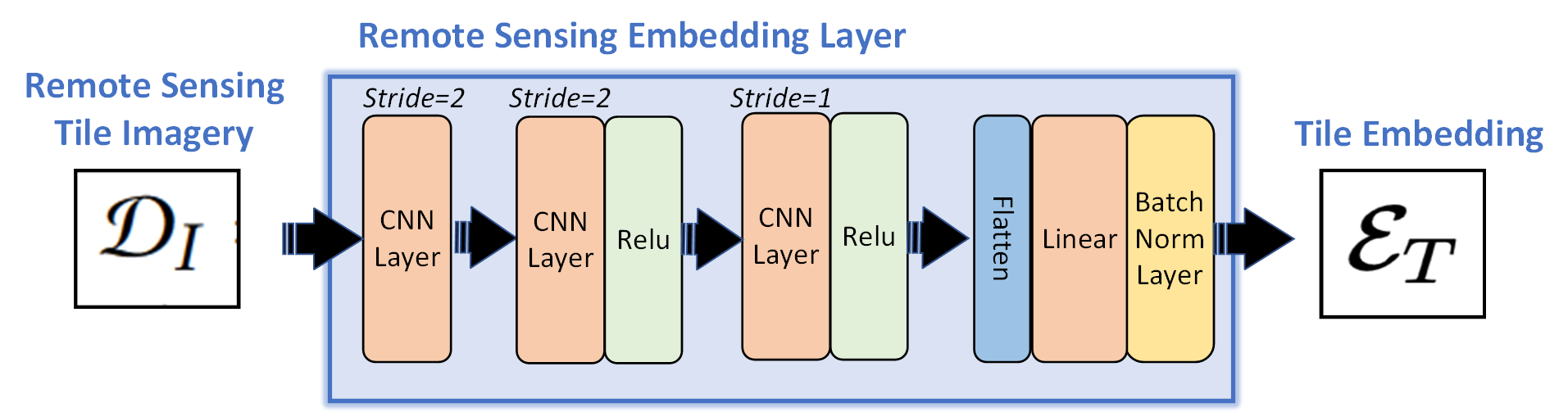}
  \vspace{-0.1in}
  \caption{The Framework of the Embedding Module $\mathcal{M} _{e_1}$.}
  \vspace{-0.15in}
  \label{fig:tile embedding}
\end{figure}

As mentioned in Section \ref{section:rs}, we leverage remote sensing images to denote tiles, so we design the image embedding module, denoted as $\mathcal{M}_{e_1}$, to covert remote sensing images $\mathcal{D}_I$ into tile embeddings $\mathcal{E}_T$. In addition, to further embed the tile sequence, corresponding to a user's given prefix trajectory, we introduce a spatial encoder and a temporal encoder to enhance tile embeddings. 

\noindent \textbf{Remote sensing image embeddings.} 
The existing research has explored various methods for image embedding. However, our current implementation faces a significant challenge due to the imperative of retaining a substantial number of generated tiles' embeddings in memory during training. We found that traditional image encoding architecture(i.e. U-Net\cite{ronneberger2015u}) struggles to handle the generation of numerous learnable embeddings, leading to high memory usage (up to 40GB). This limitation is mainly due to the substantial memory overhead from the gradient maps of these embeddings during back-propagation. Therefore, it's crucial to find a lightweight design for imagery embedding layer.

According to experimental analysis, unnecessary memory overhead mainly arises from the 2x2 max pooling layers, which retains 3/4 redundant gradients. In our model implementation, supplanting them with CNN(Convolutional Neural Network) layers using a kernel stride of two can achieve the same compression level without sacrificing effectiveness, thus saving 75\% of the memory usage. 
In particular, as depicted in \autoref{fig:tile embedding}, the final architectural configuration encompasses three successive layers of CNNs. Subsequent to this progression, the resultant compressed hyper-image, measuring 64*64, is subject to flattening prior to its traversal through a feed-forward layer. The latter serves to further compress the output features to a predetermined dimension $d_m$. Conclusively, all embeddings are uniformly normalized across the feature space, culminating in the derivation of definitive embeddings $\mathcal{E}_{T}$ corresponding to all tiles.

\begin{figure}[!t]
  \centering
  \includegraphics[width=0.9\linewidth]{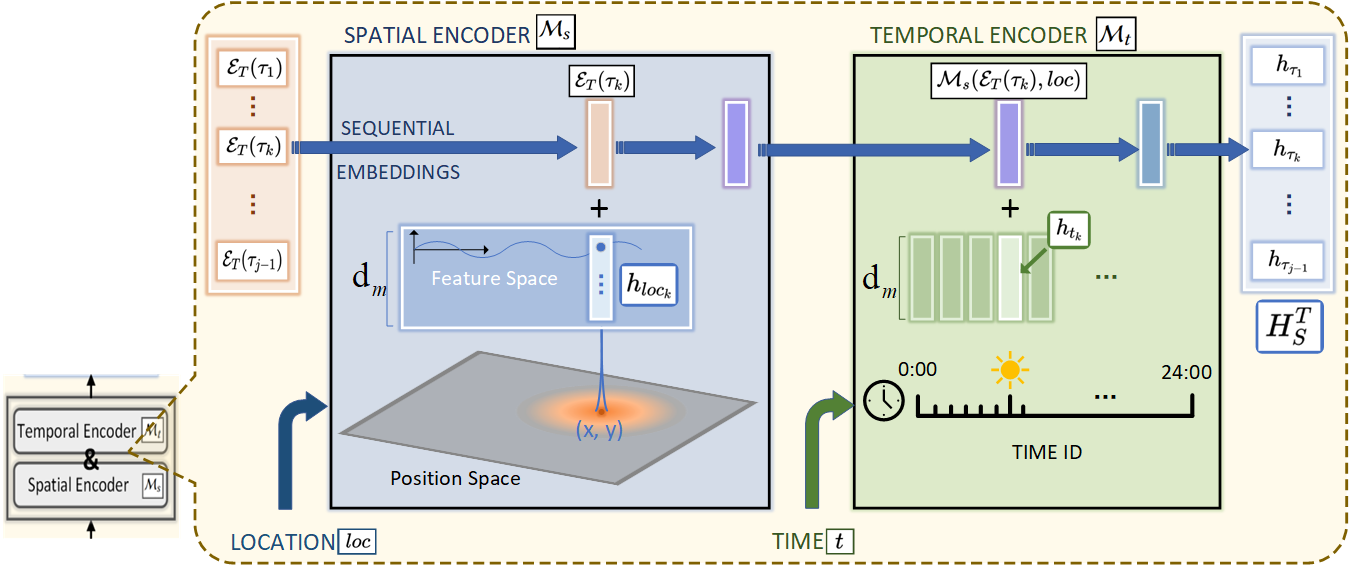}
  \vspace{-0.1in}
  \caption{The Framework of Spatial Encoder $\mathcal{M}_s$ and Temporal Encoder $\mathcal{M}_t$.}
  \vspace{-0.15in}
  \label{fig:encoders}
\end{figure}

\noindent \textbf{Enhancing with spatial encoder and temporal encoder.} As mentioned before, to capture the spatio-temporal correlation in the corresponding prefix tile sequence $[\tau_1,\cdots,\tau_{j-1}]$, each tile is augmented with a specific POI location and timestamp. To achieve this goal, we respectively devise a spatial encoder $\mathcal{M}_s$ and a temporal encoder $\mathcal{M}_t$ to enhance the embedding sequence $[\mathcal{E}_T(\tau_1), \cdots, \mathcal{E}_T(\tau_{j-1})]$. As shown in~\autoref{fig:encoders}, we first enhance the input tile sequence with locations $[loc_1,\cdots,loc_{j-1}]$ in the spatial encoder, and then further enhance the output sequence with the timestamps $[t_1,\cdots,t_{j-1}]$.

Specifically, $\mathcal{M}_s$ first encodes positional information using a sequence of sine functions with varying frequencies and phase shifts. Subsequently, these positional encodings are added to the embedding vectors, endowing them with spatial awareness. In particular, the encoding of each location $loc_k=(x_k, y_k)$ at dimension of $d_m$ is denoted as $h_{loc_k}\in \mathbb{R}^{d_m}$ in the following formulation:
\begin{equation} 
\fontsize{7.5}{10}\selectfont
h_{loc_k} \to \left \{
\begin{array}{ll}
    \hspace{1.5em}h_{loc_k}[2i]   = \sin\Big(\frac{x_k}{10000^{2i/d_m}}\Big)   &i<d_m/4\\
    h_{loc_k}[2i+1] = \cos\Big(\frac{x_k}{10000^{2i/d_m}}\Big)   &i<d_m/4\\
    \hspace{1.5em}h_{loc_k}[2i]   = \sin\Big(\frac{y_k}{10000^{2i/d_m}}\Big)   &i\ge d_m/4\\
    h_{loc_k}[2i+1] = \cos\Big(\frac{y_k}{10000^{2i/d_m}}\Big)   &i\ge d_m/4\\
\end{array}
\right.
\end{equation}

\begin{figure}[!t]
    \centering
    \includegraphics[width=0.7\linewidth]{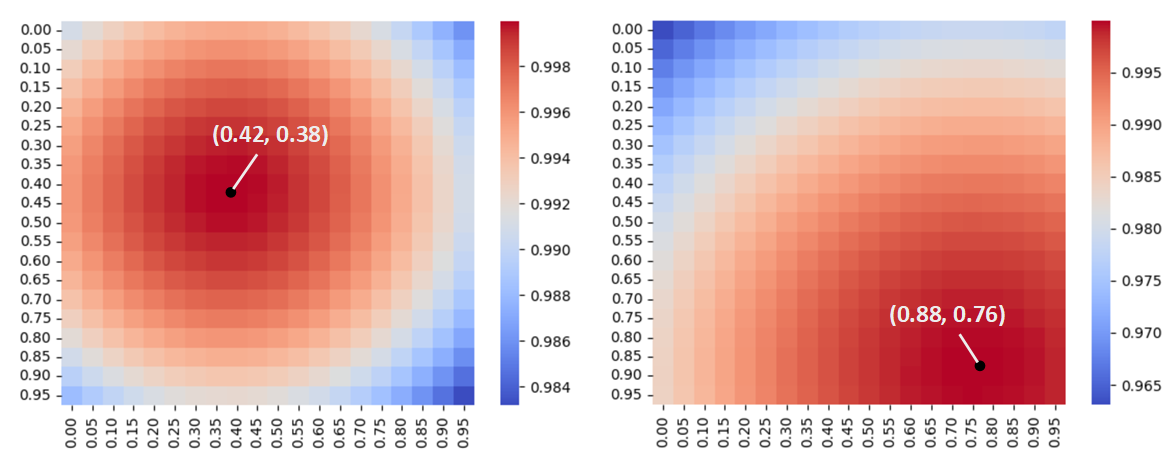}
    \vspace{-0.15in}
    \caption{These two plots show the cosine similarity of the spatial encoding between two random coordinate points (i.e., (0.42, 0.38) on the left, (0.88,0.76) on the right) and sampled points in the square space from 0 to 1.}
    \vspace{-0.15in}
    \label{fig:position similarity}
\end{figure}

Next, we can generated the corresponding spatial-enhanced tile embedding $h^s_{\tau_k}=\mathcal{E}_T(\tau_k) + h_{loc_k}$, and hence get the sequence $[h^s_{\tau_1}, \cdots, h^s_{\tau_{j-1}}]$ for the input tile sequence. Indeed, the positional encoding effectively discerns the spatial distances among distinct locations. As illustrated in~\autoref{fig:position similarity}, we graph the cosine similarity between a particular location and all other locations, leading us to deduce that heightened proximity between any two locations corresponds to greater cosine similarity in their respective positional encodings. 

Moreover, $\mathcal{M}_t$ introduces the temporal encoding to further enhance the embedding sequence. In particular, we divide a day into 48 time intervals of half an hour each. Subsequently, we map each time interval to a learnable embedding. These embeddings are then added to the embedding vectors to incorporate temporal information, and the temporal encoding process for an embedding $h^s_{\tau_k}$ can be formulated as $h_{\tau_k} = h^s_{\tau_k} + h_{t_k}$, where $h_{t_k} \in \mathbb{R}^{d_m}$ is the learnable time interval embedding. Finally, we can obtain the final tile embedding sequence, which is denoted as $H^{T}_S=[h_{\tau_1},\cdots,h_{\tau_{j-1}}]$.

\subsection{POI Embedding}
In this section, we illustrate the POI embedding module. Similar to the tile embedding module, we first design a POI information embedding module, denoted as $\mathcal{M}_{e_2}$, to provide distinguishable embeddings for each POI in set $P=\{p_1,\cdots,p_{|P|}\}$. After that, we introduce another temporal encoder to enhance POI embedding for the sample-wise prefix trajectory sequence $[(p_1,t_1),\cdots,(p_{j-1},t_{j-1})]$.

\noindent \textbf{POI information embeddings.} In POI Embedding Module $\mathcal{M}_{e_2}$, we try to utilise the extracted POI information (i.e., $id$ and $cate$) to generate corresponding POI embeddings. 
The module initially transforms the discrete data $id$ and $cate$ into continuous embedding space. Then the embeddings are aggregated according to a merging ratio $\alpha$ ($0<\alpha<1$). For POI $p$ the embedding function can be formulated as follows:
\begin{equation}
\small
\begin{aligned}
    \mathcal{E}_{P}(p)=\alpha \times embed_1(id) + (1-\alpha) \times embed_2(cate) 
    \end{aligned}
\end{equation}
where $embed(\cdot)$ is the embedding function for discrete data. 
In real-world datasets, there are tens of thousands of POI ids, with only hundreds of categories. Thus, we utilize POI $id$ to generate distinguishable embeddings, and append them with corresponding POI category embedding, which is more predictable.

\noindent \textbf{Enhancing with temporal encoder.} Similar to \autoref{section:tile}, we also enhance the POI embedding sequence $[\mathcal{E}_P(p_1), \cdots, \mathcal{E}_P(p_{j-1})]$ with timestamps $[t_1,\cdots,t_{j-1}]$. 
However, the spatial augmentation with locations $[loc_1,\cdots,loc_{j-1}]$ is not required for POI trajectory embeddings. The reason is two-fold. On the one hand, the spatial difference among different POIs can be captured from the learnable embedding of POI $id$. On the other hand, in the second prediction stage, candidate POIs are selected from predicted tiles, resulting in a high clustering of candidate POIs spatially. Hence, spatial information becomes unnecessary for POI trajectory embeddings that are leveraged in the second prediction stage. Therefore, similar to the the temporal encoding process above, a POI embedding $\mathcal{E}_{P}(p_k)$ can be augmented as $h_{p_k} = \mathcal{E}_{P}(p_k) + h_{t_k}$, where $h_{t_k} \in \mathbb{R}^{d_m}$ is the learnable time interval embedding. Finally, we can obtain the final tile embedding sequence, which is denoted as $H^{P}_S=[h_{p_1},\cdots,h_{p_{j-1}}]$.

\subsection{Historical Graph Knowledge Generation} 
In this section, we elaborate on the historical graph module $\mathcal{M}_G$, which generates tile-level and POI-level knowledge embeddings(i.e., $H^T_{\vartriangleleft}$ and $H^P_{\vartriangleleft}$) based on the given \qrpgraph $\mathcal{G}_S=\left \langle \mathcal{V}_S,\mathcal{E}_S,\Phi_S,\Psi_S \right \rangle$. In particular, we first fetch the initial embeddings from $\mathcal{E}_T$ and $\mathcal{E}_P$ for different types of nodes in $\mathcal{V}_S$. Next, considering the heterogeneous characteristic of the graph $\mathcal{G}_S$, we apply the Hetero Graph Attention(HGAT) layer to capture the correlation between different nodes. Specifically, the HGAT layer aggregates neighboring nodes' information as follows:
\begin{equation}
\fontsize{7.5}{7.5}\selectfont
\begin{aligned}
        \mathbf{A}^{l}_k[i,j] &= \text{softmax}_j(\text{LeakyReLU}(\mathbf{a}_k^l [\mathbf{W}_k^l\mathbf{h}_i^l || \mathbf{W}_k^l\mathbf{h}_j^l])) \\
        \mathbf{h}_i^{l+1} &= \sigma\left(\sum_{k\in\mathcal{R}_\omega} \sum_{j\in\mathcal{N}_k(i)}\mathbf{A}_k^l[i,j] \mathbf{W}_k^l \mathbf{h}_j^l\right)
\end{aligned}
\end{equation}
Here, $\mathcal{R}_\omega=\{Branch, Road, Contain\}$ signifies distinct categories of edges. The notation $\mathcal{N}_k(i)$ encompasses all neighboring nodes of node $i$ connected by edges of type $k$. Meanwhile, $\mathbf{A}_k^l[i,j]$ denotes the attention weight attributed to the message conveyed from node $j$ to node $i$. Node embeddings, denoted as $\mathbf{h}_i^{l}$ and $\mathbf{h}_j^l$ for nodes $i$ and $j$ respectively, play a pivotal role. The learnable attention weight for the $i$-th layer is symbolized by $\mathbf{a}_k^l$, and $\mathbf{W}_k^l$ embodies the edge weight matrix for edges of type $k$ in the $i$-th layer. To be more precise, we initially establish:
\begin{equation}
\small
h^0_i = 
\begin{cases}
    \mathcal{E}_T(i), & \text{if } \Phi(i) = tile \\
    \mathcal{E}_P(i),  & \text{if } \Phi(i) = POI
\end{cases}
\end{equation}

Subsequently, this aggregation process is iterated $n$ times to yield the ultimate embedding $h^n_i$ for each node $i$. Ultimately, we derive embeddings for both POI nodes and leaf tile nodes, culminating in $H^T_{\vartriangleleft}=\{h_i^n, \forall i \in \mathcal{V}_S \land \Phi_S(i)=tile\}$ and $H^P_{\vartriangleleft}=\{h_i^n, \forall i \in \mathcal{V}_S \land \Phi_S(i)=POI\}$.

\noindent \textit{Discussion.} Throughout the heterograph learning procedure, distinct categories of knowledge are propagated along specific types of edges. The hierarchical quad tree-based information flows via the medium of $Branch$ edges, while the interconnectivity within the road network is encapsulated by $Road$ edges. Simultaneously, the dispersion of Points of Interest (POIs) is represented by $Contain$ edges. All kinds of knowledge that interpret the correlations between user's historical trajectories and urban regions, concealed in \qrpgraph, are embedded into tile level embeddings $H^T_{\vartriangleleft}$ and POI level embeddings $H^P_{\vartriangleleft}$.

\section{Two-step Prediction}
\label{sec:5}
In this section, we aim to predict the next POI $\tau_j$ for the given embeddings: all tile embeddings $\mathcal{E}_T$,  all POI embeddings $\mathcal{E}_P$, the prefix tile sequence embedding $H^{T}_S=[h_{\tau_1}, \cdots, h_{\tau_{j-1}}]$, the prefix POI sequence embedding $H^{P}_S=[h_{p_1}, \cdots, h_{p_{j-1}}]$, and the historical tile-level and POI-level knowledge embeddings $H^T_{\vartriangleleft}, H^P_{\vartriangleleft}$. In particular, we first leverage the attention mechanism to fuse current prefix sequence embeddings and historical knowledge embedding from tile-level and POI-level views, respectively. Then, we conduct the two-step prediction procedure to generate the estimated POI $\tau_j$ based on the fused representations. 

\subsection{Attention-based Embedding Fusion}

\label{section:predictor}
As illustrated in~\autoref{fig:model architecture}, we respectively design two modules (i.e., $\mathcal{M}_{P_1}$ and $\mathcal{M}_{P_2}$) to fuse embeddings for tiles and POIs. In particular, the two modules share a similar structure with $N$ layers of attention blocks. For simplicity, we use $\mathcal{M}_{P_1}$ as the example to illustrate, where we aim to fuse the current tile sequence embedding $H^T_S$ and the historical knowledge embedding $H^T_{\vartriangleleft}$. In particular, each block in the module includes the following components:

\noindent 1. \textbf{Masked Sequential Self-attention:} To capture the sequential correlation in the sequence embedding $H^T_S=[h_{\tau_1},\cdots,h_{\tau_{j-1}}]$, we feed it into a sequential self-attention layer. Specifically, we first generate the query $Q_0^T = W_{q_0} H^{T}_S$, the key $K_0^T = W_{k_0} H^{T}_S$ and the value $V_0^T = W_{v_0} H^{T}_S$, where $W_{q_0}$, $W_{k_0}$ and $W_{v_0}$ are trainable weights. To prevent the model from utilizing information beyond the current position during attention score computation, we introduce the inverted triangle mask $M_{mask}\in \mathbb{R}^{(j-1)\times (j-1)}$, where \( M_{mask}[u,v] = {\small\begin{cases}
    1 & \text{if } u>=v \\
    -\infty & \text{if } u<v
\end{cases}} \), to mark positions after it as invalid for each given position. 
Therefore, we compute the self-attention score as $score(H^T_S, H^T_S)=softmax(\frac{Q_0^T\cdot {(K_0^T)}^{\intercal}}{\sqrt{d_m}} \cdot M_{mask}) \in \mathbb{R}^{(j-1)\times (j-1)}$. Next, the self-attention enhanced embedding sequence is computed as the weighted summation $Z_m=score(H^T_S, H^T_S) \cdot V_0^T$.

\noindent \textbf{2. Add \& Normalize:} This specific component is applied to address the issues of gradient vanishing and explosion. Specifically, we utilize the short-cut mechanism of ResNet~\cite{he2016res}. In addition, we introduce layer-normalization~\cite{lei2016ln} to mitigate the distribution discrepancy between different layers. Sequential embeddings are updated as $\overline{H}^{T}_S = LN(H^{T}_S \oplus Z_m)$, where $\oplus$ is the matrix addition operator.

\noindent \textbf{3. Cross Attention:} To take the historical knowledge embeddings $H^T_{\vartriangleleft}$ into account, we design a cross attention layer to fuse it with $\overline{H}^{T}_S$. In particular, we use $\overline{H}^{T}_S$ to compute the query  $Q_1^T = W_{q_1} \overline{H}^{T}_S$, and use $H^T_{\vartriangleleft}$ to compute the key $K_1^T = W_{k_1} H^T_{\vartriangleleft}$ and the value $V_1^T = W_{v_1} H^T_{\vartriangleleft}$. Next, we compute the score $score(\overline{H}^{T}_S, H^T_{\vartriangleleft})=softmax(\frac{Q_1^T\cdot {(K_1^T)}^{\intercal}}{\sqrt{d_m}}) \in \mathbb{R}^{(j-1)\times (|H^T_{\vartriangleleft}|)}$. At last, we get the fused embeddings $Z_h=score(\overline{H}^{T}_S, H^T_{\vartriangleleft})\cdot V_1^T$.

\noindent \textbf{4. Feed Forward:} to further enhance the fused representations, we apply a fully connected layer $Z_f=ReLU(W_fZ_h+b_f)$, where $W_f$ and $b_f$ are trainable weights and $ReLU(\cdot)$ is the activation function.

For simplicity, we denote the whole block in the $i$-layer as $AB_i(\cdot, \cdot)$, and hence obtain the final sequential output $H^{out}_{T} = AB_N ( \dots AB_2 ( AB_1 ( H^{T}_S,  H^T_{\vartriangleleft}), H^T_{\vartriangleleft} \dots), H^T_{\vartriangleleft})$. Since we aim to predict the next step, we consider the last vector in the sequential output, which is denoted as $h_{\tau}^{out}=H^{out}_{T}[-1]$. Similarly, for the POI embedding fusion module $\mathcal{M}_{P_2}$, we have the output vector $h_{p}^{out}=H^{out}_{P}[-1]$, where $H^{out}_{P}$ denotes the fused POI sequential embeddings.

\begin{figure}[!t]
  \centering
  \includegraphics[width=0.9\linewidth]{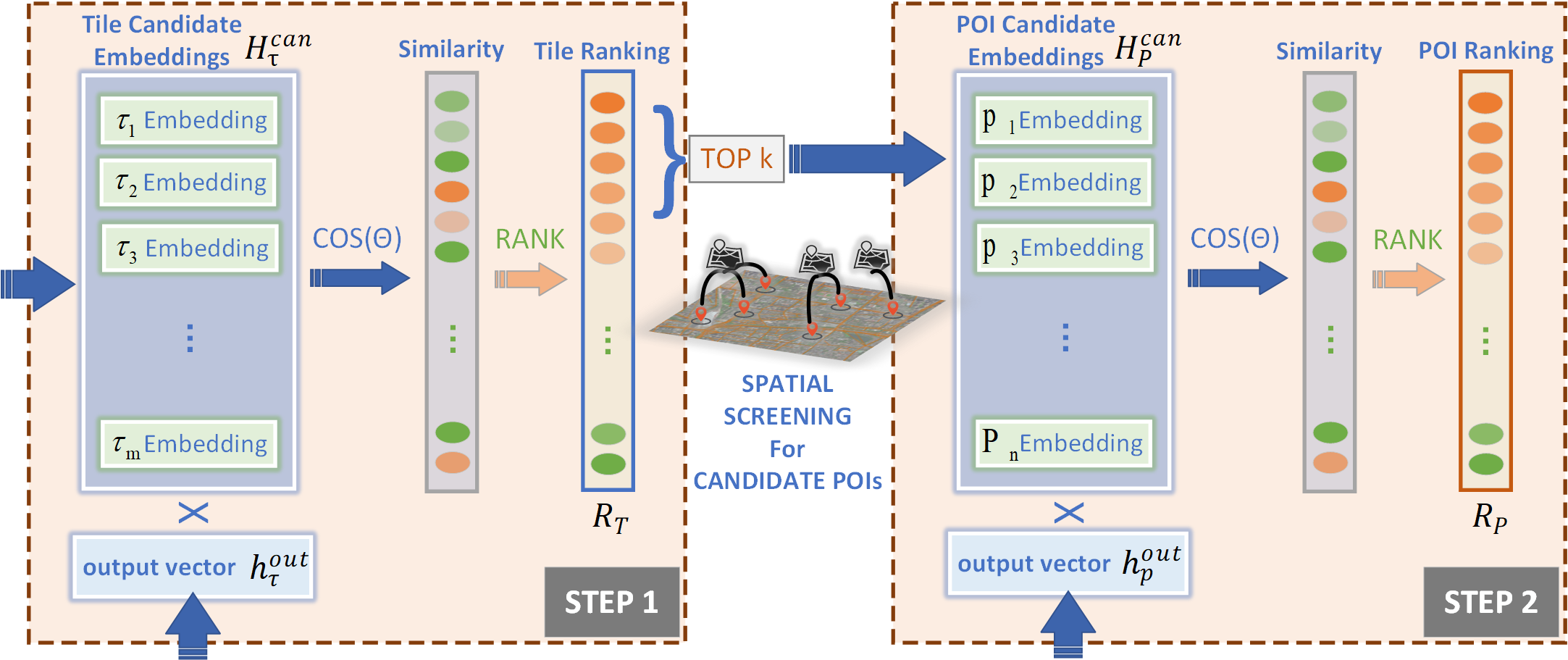}
  \vspace{-0.1in}
  \caption{An illustration of the two-step prediction process.}
  \label{fig:decision making}
  \vspace{-0.2in}
\end{figure}
\subsection{Tile Selection and POI Prediction}

As shown in~\autoref{fig:decision making}, we execute a two-step prediction process to generate the next POI prediction result. 

\noindent \textbf{Motivation.} Compared to conventional POI recommendation models\cite{liu2016predicting, feng2018deepmove, sun2020go, luo2021stan, STiSAN}, our model incorporates a tile predictor before the POI predictor, thus forming a two-step prediction process. This architecture streamlines the recommendation process by first reducing the spatial search area and then identifying targets based on user semantic preferences. By optimizing the ratio for tile selection (see \autoref{sec:two_step_interactions}), our system simplifies the task from recommending POIs out of millions to selecting specific blocks from thousands, followed by pinpointing a few POIs within those blocks. Furthermore, the tile selector doubles as a negative sample generator for the POI predictor during training, efficiently creating critical negative samples that enhance discrimination, thereby minimizing training complexity and enhancing performance.

\noindent \textbf{Tile Selection.} In the first step, all leaf tiles are selected as candidates and their corresponding embeddings are fetched to form the set $H_\tau^{can} = \{\mathcal{E}_{T}(\tau)|\mathcal{V}_\omega(\tau) = \emph{Leaf}\}$. Next, we calculate the cosine similarity between each candidate tile embedding $\mathcal{E}_T(\tau)$ and the output vector $h_{\tau}^{out}$, which is formulated as $cos(\mathcal{E}_T(\tau), h_{\tau}^{out})=\frac{{h^{out}_{\tau} \cdot \mathcal{E}_T(\tau)}}{{|h^{out}_{\tau}| \cdot |\mathcal{E}_T(\tau)|}}$. Next, all candidates are sorted descendingly by their similarities to the output vector,  forming the ranked tile candidates $R_T= [\tau_1, \cdots]$, where $cos(\mathcal{E}_T(\tau_{i_1}),h_{\tau}^{out})\geq cos(\mathcal{E}_T(\tau_{i_2}),h_{\tau}^{out})$ is satisfied for any $i_1\leq i_2$. At last, we only consider POIs within the top-$K$ tiles as input for the second step, which can significantly reduce the number of candidate POIs.

\noindent \textbf{POI prediction.} In the second step, we first select POIs with location located in the top-$K$ tiles' boundary as POI candidates, and then fetch their corresponding embeddings $H_p^{can} = \{\mathcal{E}_{P}(p)|p\rightarrow\tau, \forall \ \tau\in R_T[1:K]\}$, where the symbol "$\rightarrow$" represents that POI $p$ is located in the boundary box of tile $\tau$. At last, we also calculate and sort the cosine similarities between each embedding vector in $H_p^{can}$ and the output vector $h_{p}^{out}$, and get the sorted prediction result $R_P=[p_1, \cdots]$, where $cos(\mathcal{E}_P(p_{i_1}),h_{p}^{out})\geq cos(\mathcal{E}_P(p_{i_2}),h_{p}^{out})$ is satisfied for any $i_1\leq i_2$.

\noindent \textbf{Model Learning.} 
Actually, the actual visited POI $p_{j}$ at the subsequent time stamp $t_j$ should hold a prominent rank within the set $R_P$. This determines that, in the prediction process of our model, the next visited tile $\tau_j$ should also be highly ranked in tile ranking set $R_T$, so that $p_{j}$ can be selected as candidates from top-$K$ tiles. To achieve this, we adopted the idea from \cite{deng2019arcface} to design the following loss functions for both process:
\begin{equation}
\fontsize{7.5}{8}\selectfont
\begin{split}
    loss_\tau &= -\log\left(\frac{e^{s\cos(\theta_{\tau_j}+m)}}
    {e^{s\cos(\theta_{\tau_j} +m)}+\sum_{\mathcal{E}_{T}(\tau)\in H_\tau^{can},\tau\ne \tau_j} e^{s\cos{\theta}_{\tau}}}\right) \\
    loss_p &= -\log\left(\frac{e^{s\cos(\theta_{p_j}+m)}}
    {e^{s\cos(\theta_{p_j} +m)}+\sum_{\mathcal{E}_{P}(p)\in H_p^{can},p\ne p_j} e^{s\cos{\theta}_{p}}}\right)
\end{split}
\end{equation}

where $\theta_\tau$ is the vetorial angle between $h_{\tau}^{out}$ and tile vector $\mathcal{E}_{T}(\tau)$ (i.e., $\measuredangle( h_\tau^{out}, \mathcal{E}_{T}(\tau))$). Similarly, $\theta_p$ is the vetorial angle between $h_{p}^{out}$ and POI vector $\mathcal{E}_{P}(p)$ (i.e., $\measuredangle( h_p^{out}, \mathcal{E}_{P}(p))$). Additionally, $s$ and $m$ are hyper parameters, where $s$ is the scaling factor, while $m$ creates an angle margin between target vector and other candidate vectors. This allows model outputs (i.e., $h_\tau^{out}$ and $h_p^{out}$) to further converge towards the target vectors (i.e., $h_{\tau_j}$ and $h_{p_j}$), while pushing other candidate vectors away. Actually, the total loss is summed up by the above two losses with the following formula: $loss = \beta\times loss_\tau + loss_p$, where we use parameter $\beta$ to adjust the ratio between the two loss functions. Finally, we utilize Adam Optimizer\cite{DBLP:journals/corr/KingmaB14} to optimize all parameters by minimizing the final $loss$. This optimization process ensures that the model gradually converges towards generating a desired POI ranking list output ($R_P$).

\section{Experiments}
\label{sec:6}

\subsection{Experimental Setup\label{sec:exp_set}}
\begin{table}
\centering
{\fontsize{6.5}{7.5}\selectfont
\caption{Statistics of Datasets}
\vspace{-0.1in}
\label{tab: Dataset Statistics}
\renewcommand{\arraystretch}{1.2}
\begin{tabular}{@{\extracolsep{\fill}}c|c|c|c|c|c@{\extracolsep{\fill}}}
\Xhline{1px}
Dataset & Check-in & User & POI & Category & Coverage \\
\hline
Foursquare(NYC) & 227,428 & 1083 & 38,333 & 400 & 482.75 ${km}^2$ \\
Foursquare(TKY) & 573,703 & 2293 & 61,858 & 385 & 211.98 ${km}^2$ \\
Weeplaces(California) & 971,794 & 5250 & 99,733 & 679 & 423,967.5 ${km}^2$ \\
Weeplaces(Florida) & 136,754 & 2064 & 25,287 & 589 & 139,670.0 ${km}^2$ \\
\Xhline{1px}
\end{tabular}
}
\vspace{-0.2in}
\end{table}

\noindent \textbf{Datasets.} (1) \textbf{POI Trajectory Dataset.} 
For both training and evaluation purposes, we utilized POI check-in datasets from four distinct regions sourced from two well-known Location-Based Social Networks(LBSNs) datasets: \emph{Foursquare}\cite{yang2014modeling} and \emph{Weeplaces} \cite{xin2013personalized}. 
The Foursquare datasets respectively included check-in data in New York city(NYC) and Tokyo(TKY). The original Weeplaces dataset contained global-scale check-in data. We extracted the check-ins located in California and Florida and formed two separated datasets.
All check-in data include useful information, such as: User ID, Venue ID, Venue category ID, Location(i.e., Latitude and Longitude) and Time Stamp. 
\autoref{tab: Dataset Statistics} shows the statistics of these datasets: the Foursquare dataset and the Weeplaces dataset respectively contain trajectory data from POI sets in different spatial and sparsity configurations. The former has a large concentration of POIs in major cities, while the latter's POIs are dispersed across various regions within the state, with the latter's geographical area being approximately 1000 times larger than the former.

\noindent (2) \textbf{Remote Sensing Satellite Imagery.} 
We extract the satellite imagery sourced from \emph{Google Map} \footnote{https://www.google.com/maps} of fixed time region, using professional geographic mapping tools \emph{QGIS} and \emph{Google Earth Pro(GEP)}. For Foursquare datasets, we utilise imagery data titled "Map data ©2015 Google", covering the time period from 2014 to 2015. For Weeplace datasets, we utilise the historical google satellite imagery from GEP, covering the time period from 2010 to 2011. Moreover, we crop the remote sensing imagery based on the quad-tree structure, obtaining images with a resolution of 256 by 256 pixels and three color channels (i.e., R,G,B).

\noindent (3) \textbf{Road Network.} 
We obtain open-source road network data for the corresponding city from the public website OpenStreetMap\footnote{www.openstreetmap.org} using Python. To ensure temporal consistency with the POI Trajectory and Satellite Imagery datasets, we have integrated corresponding time period restrictions when obtaining road data.

\noindent \textbf{Evaluation Metrics.} 
We use the user's next visited position in the trajectory as ground truth for the POI prediction task.
We evaluate the performance on three kinds of metrics, namely Recall@K, NDCG@K\cite{jarvelin2002cumulated}, MRR (with $K\in\{5, 10, 20\}$). 
Specifically, Recall@K measures the hit rate of whether the top-K list contains the ground truth item, while NDCG@K, MRR evaluate the ranking accuracy according to the fact that the ground truth item should rank higher in the generated list.
In summary, higher scores for Recall@K, NDCG@K, and MRR indicate better performance.

\noindent \textbf{Baselines.} 
We compare our \mymodel\ with ten methods:
\begin{itemize}[leftmargin=10.2pt]
\small
\setlength{\itemsep}{0pt}
\setlength{\parsep}{0pt}
\setlength{\parskip}{0pt}
    \item \textbf{MC}~\cite{gambs2012next,chen2014nlpmm}: A simple probabilistic model that predicts the next item based on the current item, assuming a stationary transition probability between items.
    \item \textbf{GRU}~\cite{cho2014learning}: A variant of recursive neural network(RNN) for processing sequential data and capturing temporal dependencies. GRU applies gating mechanisms to model long-term dependencies.
    \item \textbf{STRNN}~\cite{liu2016predicting}: A reformative RNN model that extends RNN architecture with spatio-temporal features between consecutive visits.
    \item \textbf{DeepMove}~\cite{feng2018deepmove}: A state-of-the-art model which designs an attentional recurrent network that captures trajectory periodicity, and takes trajectory history into consideration. 
    \item \textbf{LSTPM}~\cite{sun2020go}: A state-of-the-art model that proposes long-term and short-term preference modeling with non-local network and geo-dilated RNN.
    \item \textbf{SAE-NAD}~\cite{SAE-NAD}: An encoder-decoder model that is consisted of a self-attentive encoder(SAE) and a neighbor-aware decoder(NAD). It differentiate user preference degrees and considered POI similarity and adjacency.
    \item \textbf{STAN}~\cite{luo2021stan}: A bi-layer attention model that incorporates spatio-temporal information and takes personalized item frequency(PIF) into consideration.
    \item \textbf{HMT-GRN}~\cite{lim2022hierarchical}: A state-of-the-art model that deals with data sparsity, by performing a Hierarchical Beam Search(HBS) with multi-task setting on spatio-temporal graph recurrent network.
    \item \textbf{Graph-Flashback}~\cite{rao2022graph}: A state-of-the-art graph model that learns representations from User-POI Knowledge Graph (STKG), and incorporate them to sequential model with defined similarity function.
    \item \textbf{STiSAN}~\cite{STiSAN}: A novel spatio-temporal interval aware model that  applies Time Aware Position Encoder (TAPE) and Interval Aware Attention Block (IAAB) enhancing sequence representations and attaching the spatial relation to self-attention.

\vspace{-.2em}
\end{itemize}

\begin{table*}[ht]
\centering
{\fontsize{7.2}{7.5}\selectfont
\caption{Result Comparison for Foursquare Dataset of Tokyo and New York}
\vspace{-0.1in}
\label{tab: foursquare result}
\centering
\scalebox{0.9}{%
\begin{tabular}{|c|*{7}{c}|}
\hline
Dataset & \multicolumn{7}{c|}{TKY / NYC}\\
\hline
\textbf{Model} & Recall@5 & Recall@10 & Recall@20 & NDCG@5 & NDCG@10 & NDCG@20 & MRR\\
\hline
MC & 0.0982 / 0.0756 & 0.1162 / 0.0807 & 0.1247 / 0.0814 & 0.0713 / 0.0561 & 0.0772 / 0.0578 & 0.0794 / 0.0580 & 0.0656 / 0.0503 \\
\hline
GRU & 0.1687 / 0.1596 & 0.2165 / 0.2076 & 0.2654 / 0.2523 & 0.1210 / 0.1131 & 0.1365 / 0.1286 & 0.1488 / 0.1399 & 0.1181 / 0.1094\\
\hline
STRNN & 0.0683 / 0.0138 & 0.0906 / 0.0231 & 0.1073 / 0.0277 & 0.0437 / 0.0095 & 0.0508 / 0.0126 & 0.0549 / 0.0138 & 0.0416 / 0.0105\\
\hline
DeepMove & 0.3035 / 0.2915 & 0.3607 / 0.3408 & 0.4081 / 0.3739 & 0.2293 / 0.2157 & 0.2479 / 0.2318 & 0.2599 / 0.2402 & 0.2185 / 0.2014\\
\hline
LSTPM & 0.3302 / 0.3041 & 0.4053 / 0.3772 & 0.4687 / 0.4362 & 0.2408 / 0.2185 & 0.2652 / 0.2422 & 0.2813 / 0.2572 & 0.2257 / 0.2069 \\
\hline
STAN & 0.1077 / 0.0792 & 0.1603 / 0.1117 & 0.2143 / 0.1495 & 0.0771 / 0.0554 & 0.0939 / 0.0659 & 0.1074 / 0.0754 & 0.0798 / 0.0562\\
\hline
SAE-NAD & 0.1575 / 0.1222 & 0.2438 / 0.2162 & 0.3629 / 0.3527 & 0.1045 / 0.0718 & 0.1323 / 0.1019 & 0.1622 / 0.1363 & 0.1067 / 0.0770 \\
\hline
HMT-GRN & 0.1537 / 0.1182 & 0.1855 / 0.1420 & 0.2115 / 0.1628 & 0.1155 / 0.0887 & 0.1258 / 0.0964 & 0.1325 / 0.1017 & 0.1107 / 0.0845\\
\hline
Graph-Flashback & 0.3094 / 0.2632 & 0.3770 / 0.3308 & 0.4375 / 0.3883 & 0.2272 / 0.1913 & 0.2491 / 0.2132 & 0.2645 / 0.2278 & 0.2154 / 0.1823 \\
\hline
STiSAN & 0.2714 / 0.2968 & 0.3322 / 0.3640 & 0.4072 / 0.4276 & 0.1877 / 0.1838 & 0.2073 / 0.2052 & 0.2261 / 0.2212 & 0.1776 / 0.1624 \\
\hlineB{2}
\mymodel & \textbf{0.3480} / \textbf{0.3095} & \textbf{0.4230} / \textbf{0.3849} & \textbf{0.4872} / \textbf{0.4487} & \textbf{0.2566} / \textbf{0.2234} & \textbf{0.2798} / \textbf{0.2483} & \textbf{0.2944} / \textbf{0.2618} & \textbf{0.2418} / \textbf{0.2117} \\
\hline
improvement & \textbf{+5.41\%} / \textbf{+1.77\%} & \textbf{+4.37\%} / \textbf{+2.04\%} & \textbf{+3.94\%} / \textbf{+2.87\%} & \textbf{+6.57\%} / \textbf{+2.24\%} & \textbf{+5.52\%} / \textbf{+2.52\%} & \textbf{+4.67\%} / \textbf{+1.79\%} & \textbf{+7.16\%} / \textbf{+2.32\%} \\
\hline
\end{tabular}}
}
\vspace{-0.2in}
\end{table*}
\begin{table*}[ht]
\centering
{\fontsize{7.2}{7.5}\selectfont
\caption{Result Comparison for Weeplaces Dataset of California and Florida}
\vspace{-0.1in}
\label{tab: weeplaces result}
\centering
\scalebox{0.9}{%
\begin{tabular}{|c|*{7}{c}|}
\hline
Dataset & \multicolumn{7}{c|}{California / Florida}\\
\hline
\textbf{Model} & Recall@5 & Recall@10 & Recall@20 & NDCG@5 & NDCG@10 & NDCG@20 & MRR\\
\hline
MC & 0.0568 / 0.0581 & 0.0635 / 0.0636 & 0.0655 / 0.0657 & 0.0436 / 0.0466 & 0.0457 / 0.0484 & 0.0463 / 0.0490 & 0.0403 / 0.0437 \\
\hline
GRU & 0.1280 / 0.1428 & 0.1619 / 0.1884 & 0.1860 / 0.2355 & 0.0943 / 0.1056 & 0.1053 / 0.1204 & 0.1114 / 0.1323 & 0.0910 / 0.1060\\
\hline
STRNN & 0.0434 / 0.0867 & 0.0782 / 0.1220 & 0.1008 / 0.1502 & 0.0217 / 0.0641 & 0.0340 / 0.0753 & 0.0491 / 0.0824 & 0.0249 / 0.0659\\
\hline
DeepMove & 0.2678 / 0.2794 & 0.3233 / 0.3514 & 0.3714 / 0.4141 & 0.2053 / 0.2140 & 0.2233 / 0.2373 & 0.2355 / 0.2533 & 0.1980 / 0.2094\\
\hline
LSTPM & 0.2971 / 0.2843 & 0.3478 / 0.3405 & 0.3916 / 0.3810 & 0.2269 / 0.2123 & 0.2434 / 0.2307 & 0.2545 / 0.2410 & 0.2135 / 0.1991 \\
\hline
STAN & 0.1326 / 0.1644 & 0.1697 / 0.2079 & 0.2089 / 0.2528 & 0.0951 / 0.1119 & 0.1070 / 0.1262 & 0.1169 / 0.1375 & 0.0917 / 0.1057\\
\hline
SAE-NAD & 0.0820 / 0.1142 & 0.1468 / 0.2120 & 0.2602 / 0.3210 & 0.0496 / 0.0669 & 0.0703 / 0.0985 & 0.0988 / 0.1261 & 0.0551 / 0.0722 \\
\hline
HMT-GRN & 0.1978 / 0.1966 & 0.2201 / 0.2273 & 0.2343 / 0.2364 & 0.1535 / 0.1475 & 0.1607 / 0.1575 & 0.1643 / 0.1599 & 0.1433 / 0.1379\\
\hline
Graph-Flashback & 0.2324 / 0.2690 & 0.2850 / 0.3432 & 0.3338 / 0.4200 & 0.1744 / 0.1990 & 0.1915 / 0.2231 & 0.2038 / 0.2423 & 0.1674 / 0.1938 \\
\hline
STiSAN & 0.1894 / 0.2273 & 0.2292 / 0.2614 & 0.2746 / 0.2841 & 0.1318 / 0.1337 & 0.1448 / 0.1451 & 0.1565 / 0.1506 & 0.1239 / 0.1146 \\
\hlineB{2}
\mymodel & \textbf{0.3083} / \textbf{0.3057} & \textbf{0.3587} / \textbf{0.3676} & \textbf{0.4063} / \textbf{0.4227} & \textbf{0.2383} / \textbf{0.2319} & \textbf{0.2550} / \textbf{0.2519} & \textbf{0.2669} / \textbf{0.2653} & \textbf{0.2286} / \textbf{0.2236} \\
\hline
improvement & \textbf{+3.77\%} / \textbf{+7.53\%} & \textbf{+3.13\%} / \textbf{+4.61\%} & \textbf{+3.75\%} / \textbf{+0.64\%} & \textbf{+5.02\%} / \textbf{+8.36\%} & \textbf{+4.77\%} / \textbf{+6.15\%} & \textbf{+4.87\%} / \textbf{+4.74\%} & \textbf{+7.07\%} / \textbf{+6.78\%} \\
\hline
\end{tabular}}
}
\vspace{-0.2in}
\end{table*}

\noindent \textbf{Implementation Details.}

We first divide user visited POI sequence into trajectories. Then we randomly select 80\% of the trajectory dataset as the training set, 10\% as the validation set, and 10\% as the test set. During training, we use the Adam optimizer and the common hyper-parameters are set as follows: 
\emph{batch\_size = 8, learning\_rate = 2e-5(with 0.95 decay), dropout = 0.1, epoch = 40. }
The \qrpgraph settings(i.e. \emph{D}: maximum tree height, $\Omega$: The maximum number of POI in leaf tiles) and parameter $K$(i.e. top $K$ tiles) are correlated with the the dataset specifics. We set {\small\{\emph{D}=8, $\Omega$=100, $K$=15\}} for Foursquare\_TKY, {\small\{\emph{D}=8, $\Omega$=50, $K$=15\}} for Foursquare\_NYC, {\small\{\emph{D}=9, $\Omega$=100, $K$=10\}} for Weeplaces\_California, {\small\{\emph{D}=8, $\Omega$=50, $K$=10\}} for Weeplaces\_Florida.

For all baselines we follow their default recommended settings, while made some adjustment to the settings for new datasets if needed. {\small \emph{Remark: We did not follow some baselines (\cite{luo2021stan,lim2022hierarchical}) to intentionally filter out less frequent POIs, since it's unreasonable to foresee POI frequency in real-world scenarios.}}

\begin{table}
\caption{Results of Ablation Experiments}
\vspace{-0.1in}
\label{tab: Ablation result}
{\fontsize{6.5}{7.5}\selectfont
\centering
\setlength\tabcolsep{2pt} 
\begin{tabular}{|c|c|c|c|c|c|}
\hlineB{2}
\multicolumn{2}{|c|}{Dataset} & \multicolumn{4}{c|}{TKY / NYC}\\
\hlineB{2}
\multicolumn{2}{|c|}{Model} & Recall@5 & NDCG@5 & MRR & impro@avg\\
\hline
\multicolumn{2}{|c|}{\textbf{\mymodel}} & \textbf{0.3480} / \textbf{0.3095} & \textbf{0.2566} / \textbf{0.2234} & \textbf{0.2418} / \textbf{0.2117} & - / - \\
\hline
\multicolumn{2}{|c|}{\makecell{Grid Replace \\Quad-tree}} & 0.2673 / 0.2416 & 0.2047 / 0.1744 & 0.1937 / 0.1637 & -21.10\% / -22.18\% \\
\hline
\multicolumn{2}{|c|}{No Two-step} & 0.0454 / 0.0435 & 0.0319 / 0.0271 & 0.0382 / 0.0355 & -86.24\% / -85.68\%\\
\hline
\multirow{3}{*}{\tiny QR-P}
 & No Graph & 0.0815 / 0.1176 & 0.0654 / 0.0898 & 0.0631 / 0.0875 & -75.00\% / -60.16\% \\
 & No Contain & 0.2746 / 0.2401 & 0.1999 / 0.1731 & 0.1902 / 0.1655 & -21.51\% / -22.25\%\\
 & No Road & 0.3357 / 0.2667 & 0.2443 / 0.1910 & 0.2298 / 0.1811 & -4.43\% / -14.26\%\\
\hline
\multicolumn{2}{|c|}{No Imagery} & 0.3062 / 0.2839 & 0.2209 / 0.2028 & 0.2090 / 0.1911 & -13.16\% / -9.07\%\\
\hline
\multicolumn{2}{|c|}{No S\&T Encoder} & 0.3218 / 0.2825 & 0.2309 / 0.1991 & 0.2171 / 0.1867 & -9.25\% / -10.47\%\\
\hline
\multicolumn{2}{|c|}{No POI Category} & 0.2780 / 0.2577 & 0.2035 / 0.1831 & 0.1897 / 0.1751 & -20.79\% / -17.35\%\\
\hlineB{2}
\end{tabular}\\
\footnotesize  \quad(impro@avg: average rate of improvements)
}
\vspace{-0.3in}
\end{table}

\subsection{Effectiveness Evaluation}
We compare the performance of different models on two urban datasets and two state datasets based on three kinds of metrics. We report the average value of five experiments with different random seeds to ensure the validity of the evaluation.

\noindent \textbf{Performance Overview.} As illustrated in \autoref{tab: foursquare result}\&\autoref{tab: weeplaces result}, our model consistently outperforms all baseline models across all metrics. In all, our model achieve an average improvement of 5.4\% in TKY, 2.2\% in NYC, 4.6\% in California and 5.5\% in Florida across all metrics. In particular, our model performs exceptionally well on the Florida dataset, with an average NDCG improvement of 6.4\%. Additionally, several metrics on other datasets also show improvements exceeding 6\%. Even the metric with the smallest improvement across all tests remains on par with the best baseline.

\noindent \textbf{Comparison Analysis.} According to the results, we have the following observations:

\noindent (1) Traditional non-deep learning methods, like MC, are characterized by their simplistic nature, as they acquire prediction ranking values through predefined and unchanging methods, leading to suboptimal performance. This outcome effectively substantiates the compelling rationale behind the utilization of deep learning models in this predictive domain.

\noindent (2) To evaluate the effectiveness of recurrent neural networks, we present the results for GRU as baseline. It is evident that RNN model exhibit superior performance in comparison to non-deep-learning techniques. However, its model structure is too simplistic and may not be suitable for direct application on this tasks.

\noindent (3) In the case of HMT-GRN, a model employing a tile selection strategy akin to ours, it is evident that its applicability is primarily confined to datasets characterized by globally-scaled POI distributions. Notably, even after adapting their tile encoding scheme to the urban context, their beam search mechanism struggles to effectively discriminate the target POI from alternative options. 

\noindent (4) For STiSAN, it performs fairly well on urban datasets but has mediocre results on state-level datasets. This is because during training, it can only specify the nearest POIs beside the target as negative samples. Therefore, when tested datasets with sparse POI distributions may pose challenges to its predictions.

\noindent (5) The most competitive baselines for our model are DeepMove, LSTPM, Graph-Flashback and STiSAN. Both DeepMove and LSTPM take historical trajectories into consideration, and have specialized modules for handling long-term historical trajectories, making them suitable for trajectories that have non-continuous intentions. In the case of Graph-Flashback, its methodology revolves around the incorporation of trajectory information through the establishment of a knowledge graph structure. This straightforward yet impactful approach facilitates the acquisition of historical trajectory knowledge, eliminating the requirement for consecutive POI records. 
In contrast, alternative models like STRNN and STAN primarily rely on transition matrices for prediction, falling short in their predictive efficacy. SAE-NAD considered user historical trajectory as a check-in set, causing the predicted embedding not effective for every types of sequential combination.
These models exhibit subpar performance across both datasets, underscoring their limited generalization capabilities.

\subsection{Ablation Study}

To get a better view of the contribution of different modules, we have to conduct ablation experiments. In particular, we propose four variations of our model, which are defined as follows:

\begin{itemize}[leftmargin=10.2pt]
\small
\setlength{\itemsep}{0pt}
\setlength{\parsep}{0pt}
\setlength{\parskip}{0pt}
    \item \emph{Grid Replace Quad-tree}: We replace \qtree with a grid of fixed granularity. (We test multiple granularity of grids and report the best result.)
    \item \emph{No Two-step}: To verify the two-step is a vital structure, we bypass the tile selector and directly allow POI predictor to make the decision.
    \item \emph{No QR-P graph}: We remove the \qrpgraph input along with the \textit{Historical Graph Knowledge} encoding module.
    \item \emph{QR-P with no Road}: Fine-grained ablation. we remove the "Road" edges from \qrpgraph input for fine-grained ablation.
    \item \emph{QR-P with no Contain}: We remove the "Contain" edges from \qrpgraph input for fine-grained ablation.
    \item \emph{No Remote Sensing}: We remove the Remote Sensing Imagery input along with its embedding module. 
    \item \emph{No S\&T Encoder}: We remove all the Spatial and Temporal Encoder layers from the model architecture.
    \item \emph{No POI Category}: To prove category is useful for POI embeddings, we remove the category information to let POI embeddings generated only with their ids.
\vspace{-.2em}
\end{itemize}

We report the result of ablation experiments in \autoref{tab: Ablation result}. According to the results, we have the following observations: 
\noindent (1) Replacing quad-tree with a fixed-granularity grid results in a significant performance decrease of more than 20\% for the model. This also indicates that a \qtree-based multi-granularity partitioning of space indeed enhances spatial representation capabilities and spatial-level filtering effectiveness for POI.

\noindent (2) Bypassing the tile selection process would result in a significant decline in performance, making the task of POI prediction overly burdensome and unable to provide effective recommendations. This demonstrates the critical importance of the tile filtering process.

\noindent (3) The \qrpgraph and the incorporation of Historical Graph Knowledge emerge as the primary drivers behind \mymodel's effectiveness. Without this integral module, \mymodel\ would experience substantial deterioration across all metrics, with more than a 60\% decline from the original values. Moreover, fine-grained ablations prove the \emph{"Road"} and \emph{"contain"} edges can significantly improve the graph's effectiveness.

\noindent (4) The integration of \textit{Remote Sensing}, \textit{S\&T Encoders}, and \textit{POI Category} contributes to the augmentation of embeddings. Notably, the \textit{POI Category} exerts the most profound influence on model performance, enhancing the metrics by an average of 17.35\% and 20.79\% for each datasets. Both \textit{Remote Sensing} and \textit{S\&T Encoder} bring about average improvements ranging from 9\% to 13\%.

\begin{table}
\centering
\caption{Model efficiency comparison}
\vspace{-0.1in}
\footnotesize
\label{tab: Efficiency}
{\fontsize{8}{8}\selectfont
\centering
\scalebox{0.8}{
\begin{tabular}{|c|*{3}{c}|*{3}{c}|}
\hline
Dataset & \multicolumn{3}{c|}{NYC} & \multicolumn{3}{c|}{TKY}\\
\hline
\textbf{Model} & Memory & Train & Infer & Memory & Train & Infer\\
\hline
\mymodel & 14,111M & 03:41 & 02:20 & 15,667M & 11:21 & 07:23 \\
\hline
STAN & 13,517M & 141:39 & 08:25 & 26,271M & 531:59 & 42:22 \\
\hline
HMT-GRN & 10,189M & 08:49 & 29:12 & 10,379M & 32:26 & 110:08 \\
\hline
DeepMove & 2,799M & 18:53 & 07:27 & 3,723M & 71:28 & 25:32 \\
\hline
LSTPM & 9,334M & 04:03 & 02:35 & 19,197M & 12:05 & 09:29 \\
\hline
Graph-Flashback & 3,159M & 0:14 & 01:19 & 3,776M & 0:18 & 02:45 \\
\hline
STiSAN & 14,353M & 06:55 & 01:19 & 14,385M & 28:36 & 07:09 \\
\hline
\end{tabular}}\\
\fontsize{7.5}{8}\selectfont \quad(M - mega byte, num:num - minute:second)
}
\vspace{-0.1in}
\end{table}

\subsection{Efficiency Evaluation}

We report the efficiency of our \mymodel\ comparing with some effective baselines in this section. 
As shown in \autoref{tab: Efficiency}, we report the memory cost(denoted as \emph{Memory}), training time(denoted as \emph{Train}), and the inference time(denoted as \emph{Infer}) of all models on two datasets. The results demonstrate that our model ranks among the top on inference efficiency, and it also has shorter training time compared to all models except Graph-Flashback. Graph-Flashback achieves a high training speed by utilizing a batch size of 200 to 1024, but it requires 100 epochs to converge.
Moreover, though our model exhibits a higher memory cost on the NYC dataset than some other models due to our use of multimodal data, it still significantly outperforms STAN and LSTPM in terms of memory consumption on the larger TKY dataset. 


\subsection{Parameter Analysis}

\begin{figure}
  \centering
  \includegraphics[width=0.98\linewidth]{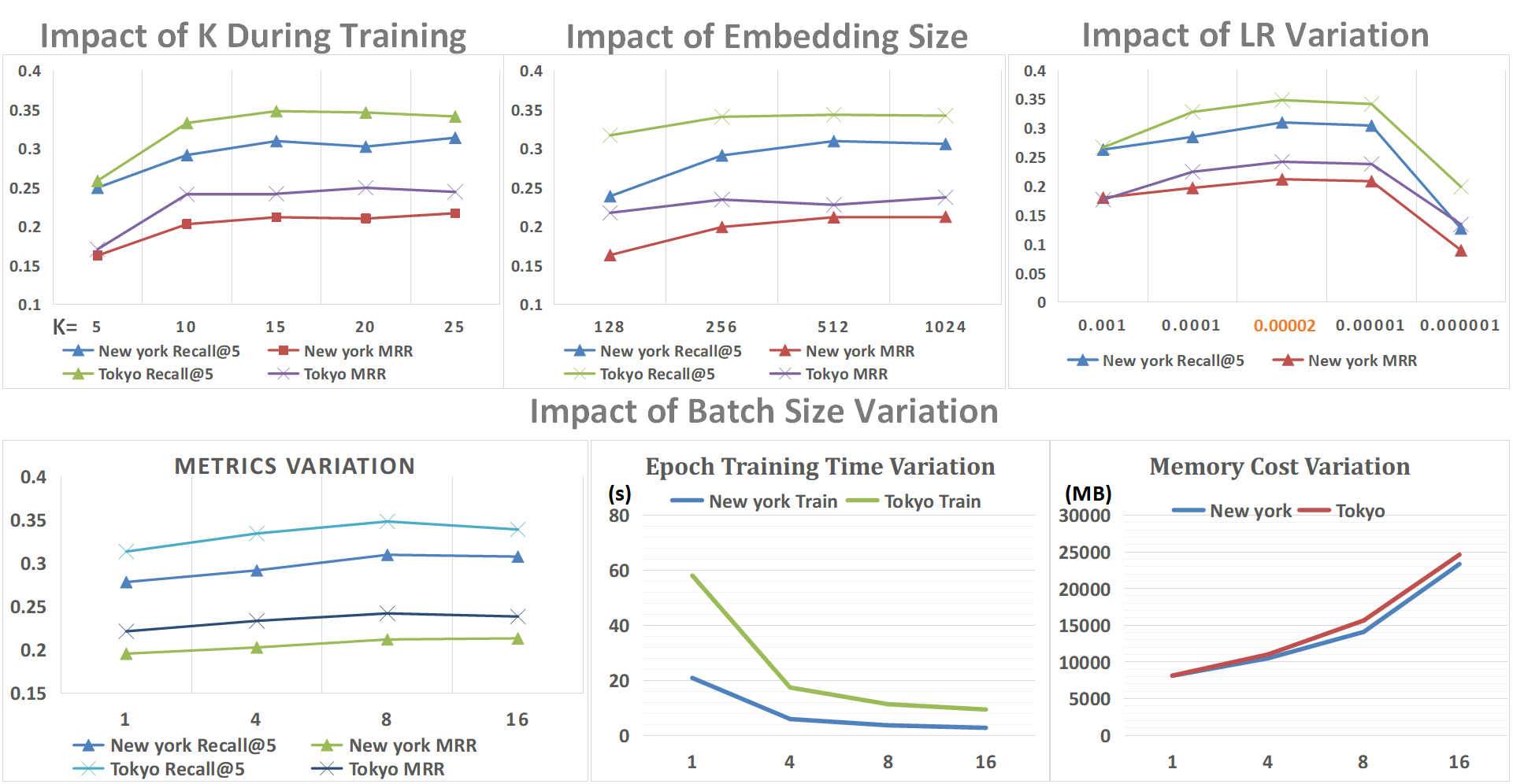}
  \vspace{-0.1in}
  \caption{Evaluation of Parameter Tuning Performance.}
  \vspace{-0.25in}
  \label{fig:hyperparam 2}
  
\end{figure}

As shown in \autoref{fig:hyperparam 2}, we evaluate the performance and impact of three parameters' tuning (i.e., embedding dimension $d_m$, learning rate $lr$ and the batch size) on two datasets. Metrics $Recall@5$ and $MRR$ are selected to represent the accuracy and ranking performance.

\noindent \textbf{Param K}(during training)\textbf{.} We sample $K$'s value from \{5, 10, 15, 20, 25\}. When $K$ exceeds 10, the metrics for evaluation exhibit less than a 5\% fluctuation. However, when $K$ is set below 10, a significant drop in performance is observed. This is because setting K too small during training will result in too few negative samples in the POI candidate set, making accurate training difficult. Hence, setting K to a certain range above 10 is all acceptable.

\noindent \textbf{Embedding Dimension.} We vary the dimension of embeddings $d_m$ for the embedding modules $\mathcal{M}_{e_1}$,$\mathcal{M}_{e_2}$ from 128 to 1024 with scale of 2. The result shows that embedding dimension of 512 is the best size for dataset NYC, while the dimension size doesn't seem to have much impact when training dataset TKY. Considering the generality, we set 512 as the embedding dimension for \mymodel.

\noindent \textbf{Learning Rate.} We examine various learning rates denoted as $lr$, ranging from 0.000001 to 0.001. The upper-right line chart depicted in \autoref{fig:hyperparam 2} shows the main variation along with the convergence epochs, where the learning rate around 0.00002 has the optimal performance. Therefore, the utilization of a learning rate set at 0.00002 in terms of both convergence velocity and comprehensive performance.

\noindent \textbf{Batch Size.} 
We assessed the impact of different batch sizes on the second line of \autoref{fig:hyperparam 2}. The revealed slight distinctions in the evaluation results affirms the stability of our model. It is also noteworthy that the alterations in training time and memory requisites follow an exponential pattern. In particular, training with batch size of 1 leads to the longest training duration with the least memory cost, while training with batch size of 16 shows the opposite outcome. Therefore, a batch size of 8 is selected as the default value due to its optimal performance across all metrics.

\subsection{Interactions Between the Two Steps.}
\label{sec:two_step_interactions}
\begin{figure}
  \centering
  \includegraphics[width=0.9\linewidth]{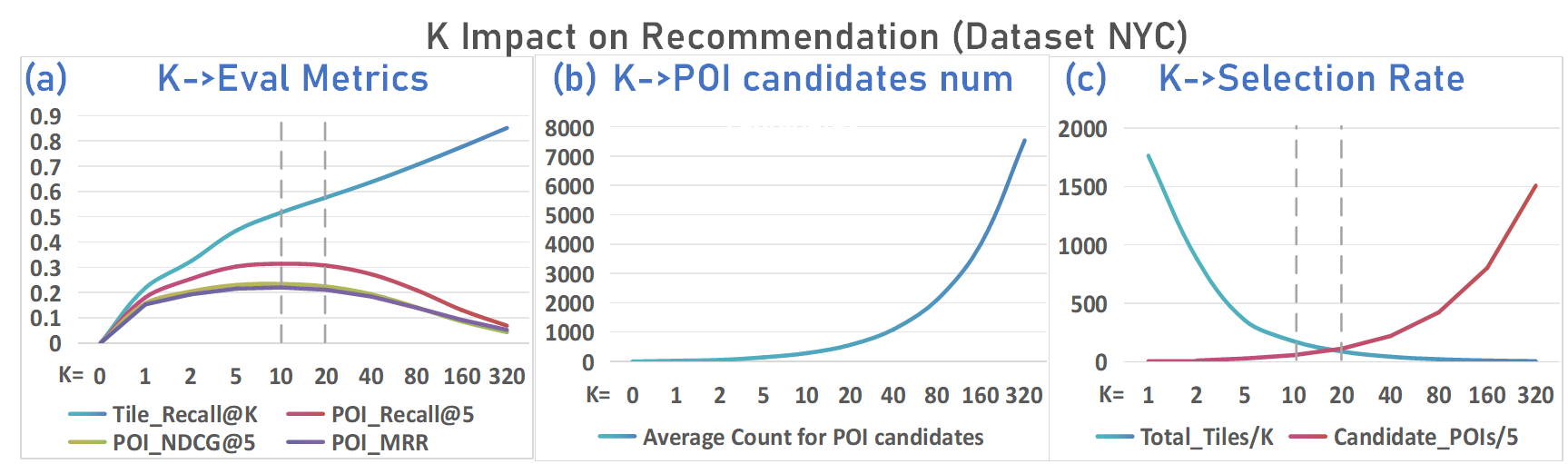}
  \vspace{-0.05in}
  {\fontsize{6.5}{8}\selectfont
  \quad(Selection Rate: A/B indicates the difficulty of selecting B items out of A items)}
  \vspace{-0.05in}
  \caption{Impact of top $K$ tiles.}
  \vspace{-0.2in}
  \label{fig:hyperparam 1}
  
\end{figure}

To illustrate how the two-step architecture interacts with each other during inference, we sampled $K$ from 1 to 320 at intervals of 2 exponential steps, and tested the trained model on the NYC dataset, the statistical results are shown in \autoref{fig:hyperparam 1}. In (a), it is evident that the accuracy of the top $K$ tiles unquestionably improves as $K$ increases. However, the accuracy of the top 5 POIs stopped increasing after reaching its peak at around K=10 to 20. This is mainly because the increase in $K$ leads to a decrease in the ability to narrow down the search space in the first step, thereby increasing the burden of the second step of POI prediction. As seen in (b), the candidate set of POIs in the second step increases exponentially with the growth of K.
The difficulty variations of the two-step tasks, as depicted in (c) through the selection rate, illustrate that the difficulty of predicting the top $K$ tiles and predicting the top 5 POIs respectively decrease and increase exponentially with the growth of K. The intersection of their curves precisely aligns with the peak accuracy position of the top 5 POIs, at which point the difficulty of both steps remains relatively low.

\subsection{Case Study}

\begin{figure}[!t]
  \centering
  \includegraphics[width=0.92\linewidth]{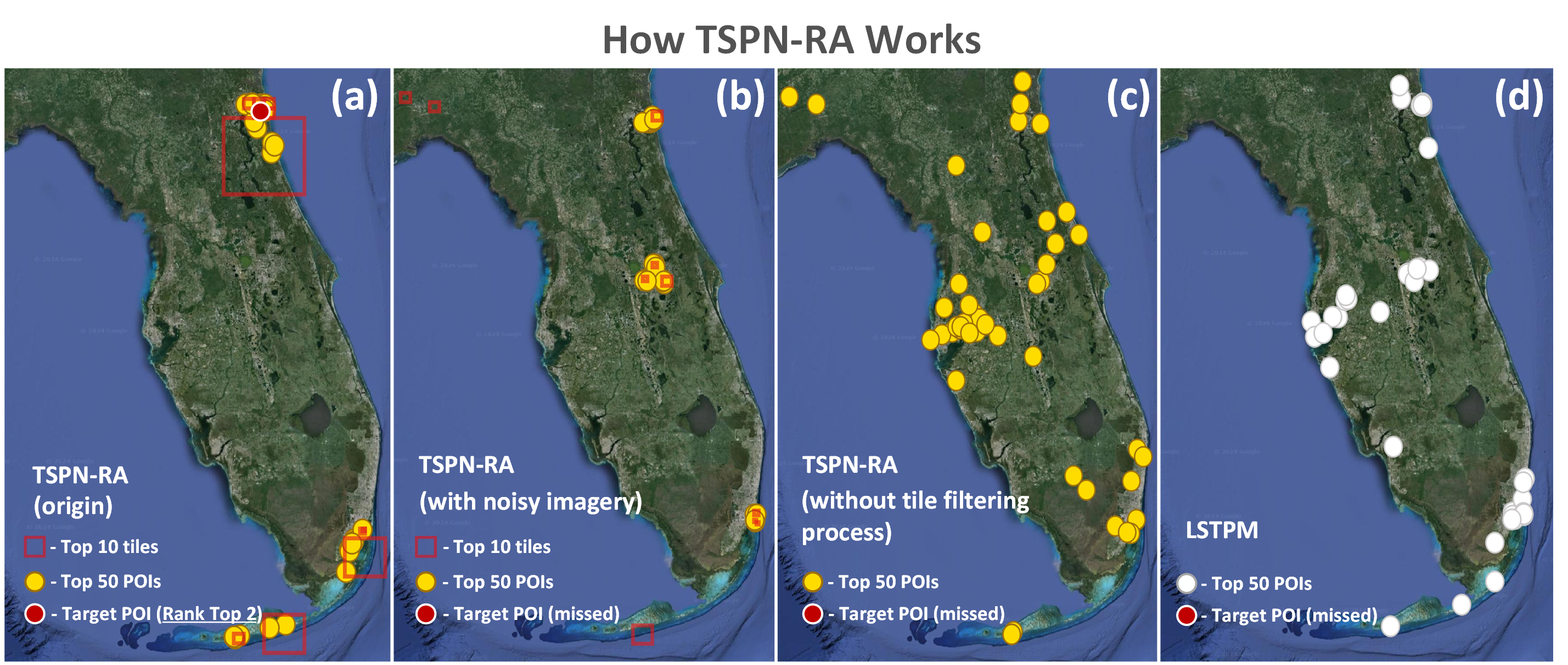}
  \vspace{-0.1in}
  \caption{A comparison of models' recommendations on a case in Florida. A user who was active on eastern coast was trying to visit a POI named Beswick-island(target POI) in Jacksonville.}
  \vspace{-0.2in}
  \label{fig:CaseStudy}
\end{figure}

To visually illustrate the advantages of our model, we conducted a case study to qualitatively demonstrate its performance. We extracted a trajectory of a user who was active on the eastern coast of Florida, with the target destination being a POI in Jacksonville. As depicted in \autoref{fig:CaseStudy}, we compared the distribution of the top 50 recommended POIs provided by four methods: (a)\mymodel, (b)\mymodel(with noisy imagery), (c)\mymodel(without tile filtering process), and (d)the best-performing baseline \textbf{LSTPM}. According to the result, we have the following observations:

(1) \autoref{fig:CaseStudy}(a) shows that the top 50 POIs recommended by \mymodel\ are all distributed along the eastern coastline of Florida, aligning perfectly with the characteristics of the target POI location. In contrast, \autoref{fig:CaseStudy}(d) shows that the baseline lacks environment contextual understanding, resulting in the top 50 POI recommendations being randomly distributed across POI densely populated areas.

(2) To evaluate the satellite imagery's effectiveness, we introduced 20\% noise to the imagery data, and the result is depicted in \autoref{fig:CaseStudy}(b). It can be observed that the majority of tiles now appear in inland areas, resulting in failed predictions. This demonstrates the significance of satellite imagery in capturing the feature of the `eastern coastline' and aiding the spatial filtering process of POIs.

(3) To evaluate the two-step framework's effectiveness, we bypassed the tile selection and directly allowed the POI predictor to make the decision. As shown in \autoref{fig:CaseStudy}(c), the model missed the target, and the distribution of the top 50 POIs scattered across various regions. This confirms that providing results directly from the entire POI dataset can lead to confusion in certain scenarios, whereas our model's spatial-semantic separation-based prediction indeed offers positive spatial filtering for POIs.

\section{related work}
\label{sec:7}

\noindent \textbf{Spatio-temporal POI Prediction.}
Next POI prediction is an important topic oriented from POI recommendation system, various approaches have been explored to tackle this spatio-temporal prediction task. 
Traditional methods, such as Markov-based models, estimate the next behavior probability using transition matrices\cite{koren2009matrix}. Varietal models such as \cite{rendle2010factorizing, cheng2013you} learn personalized transition matrix to produce user with personalized recommendation. However, these models primarily focus on the transition probability between consecutive visits, making them susceptible to the sparsity of POI sequential data.

To cope with the unresolved challenge, researchers have turned to study deep learning models for resolutions. 
RNN-based models are the most commonly used. The basic RNN model and its variants can be used directly for the task, but with low accuracy and robustness. By extending RNN, Liu et al.\cite{liu2016predicting} introduced spatial temporal information with distance and time transition matrices. Some works such as Sun et al.\cite{sun2020go} focus on learning Long Short-Term preference leveraging the advantage of LSTM. 
Attention-based methods have also been widely researched. The work of Feng et al.\cite{feng2018deepmove} can be considered a successful endeavor in the design of an attention-based recurrent neural network model. Subsequently, other attention models\cite{SAE-NAD, luo2021stan, lim2021origin, li2021mgsan, wang2021adq, STiSAN, DBLP:conf/icde/00020BF21,DBLP:conf/sigmod/Yuan0BF20,DBLP:conf/www/LiYFMXW23,DBLP:journals/pvldb/00020B22,DBLP:journals/pvldb/0002WBW23} focus on incorporate knowledge from different aspects.  
In recent works, graph-based approaches have also become more prevalent. Rao et al.\cite{rao2022graph} applies a simplified Graph Convolution Network(GCN) on the POI transition graph to enrich the POI representations. Lim et al.\cite{lim2022hierarchical} utilizes a Graph Recurrent Network and introduces a Hierarchical Beam Search method to predict POI with multiple steps.
However, all the previous approaches are poor in dealing with challenges we mentioned earlier.

\noindent \textbf{Deep Learning with Spatio-Temporal Data.}
With the increasing adoption of deep learning models, a growing volume of spatio-temporal data is being leveraged. 
\emph{Recurrent Neural Networks (RNNs)} have been recently deployed for trajectory modeling. For instance, Wu et al.\cite{DBLP:conf/ijcai/WuCSZW17} utilize RNNs to predict future movements, outperforming traditional models. Similarly, in \cite{ihmvte}, authors embed trajectories in an RNN model to identify user mobility patterns. 
\emph{Convolutional Neural Networks (CNNs)} have also received considerable attention. Song et al.\cite{pshmtmcl} propose an intelligent transportation system that simulates human mobility and transportation modes using CNNs. In \cite{dbmfsd}, the study treated crowd density as images and further proposed CNNs-based network for predicting crowd flows.
\emph{Graph neural networks} further considers the road network's graph structure to tackle traffic prediction problems. Examples include travel demand prediction \cite{igaracwtdp} and traffic flow prediction \cite{utpstdudml}.
\emph{The attention mechanism} has been applied to capture complex spatio-temporal correlations among different features. For example, Yuan et al.\cite{aejpmtdtf} employ attention mechanisms to jointly predict travel demands and traffic flows by capturing diverse spatio-temporal correlations.
Recent studies \cite{alex, PGM, FIT, MADEX, ASOTP} explore methods for designing adaptive learned indexes to support data updating. Other works \cite{MLindex, lmdi, tclsi, Qd-tree} investigate how to build learned indexes specifically for multi-dimensional data, which are more applicable.

\section{conclusion}
\label{sec:8}
Our study proposed a two-step POI recommendation model from separated spatial and semantic perspectives. We developed an improved graph construction method for capturing the semantic and spatial relationships within historical trajectories, forming the QR-P graph. Additionally, we incorporated remote sensing data to augment region representation in the spatial dimension. Extensive experiments showed that our methods outperformed state-of-the-art by 2\%-8\% on four real-world datasets with different spatial sparsity. 
Our study advances intelligent city mapping by introducing a novel remote sensing embedding approach and offers valuable insights into urban POI recommendations. In the future, we will explore additional applications of remote sensing data augmentation in the intelligent cities research field.

\newpage
\section{Acknowledgments}
This work is supported by the National Natural Science Foundation of China under Grant No. 62032003 and Grant No. 61921003. 
\bibliographystyle{IEEEtran}
\bibliography{reference}

\end{document}